# DiffSIM: Unconditional and conditional facies simulation based on denoising diffusion generative models

**Minghui Xu[1], Suihong Song[1], Tapan Mukerji[1,2,3]**

[1]Department of Energy Science & Engineering, Stanford University, Stanford, CA 94305, USA

[2]Department of Geophysics, Stanford University, Stanford, CA 94305, USA

[3]Department of Earth & Planetary Sciences, Stanford University, Stanford, CA 94305, USA

Key Points:

- Developed two diffusion-based generative models for facies-model simulation
- Demonstrated high-fidelity unconditional generation using multiple distributional metrics
- Introduced a mask-based conditioning strategy to honor hard well facies constraints

**Abstract**

Constructing subsurface facies models that are geologically plausible and constrained by well facies is essential for analyzing sedimentary evolution, reservoir characterization, and flow simulation. Recent deep generative model-based geomodelling methods have demonstrated promising capabilities for both unconditional and conditional settings. We investigate denoising diffusion models as a generative framework for producing realistic and diverse facies realizations in both settings. Diffusion models generate samples through a fixed forward noising process and a learned reverse denoising process. For unconditional geomodelling, we use denoising diffusion probabilistic models (DDPMs) to learn geological patterns from training facies models, and adopt denoising diffusion implicit models (DDIMs) to accelerate sampling by reducing inference steps from 1500 to 50 (30× fewer steps). We assess the geological plausibility using data distribution, class proportions, variograms, and geometric features. Quantitative comparisons show that the generated realizations closely match the test set across these metrics, indicating that diffusion models capture realistic geological patterns while maintaining diversity. To enable conditional generation, we encode well facies and their spatial locations as conditional indicators and apply a mask-based denoising strategy that generates facies only in between-well regions, guaranteeing hard conditioning without introducing additional loss weights. We evaluate both unconditional and conditional generation on three scenarios: two two-dimensional cases (meandering channels and point bars) and one three-dimensional point-bar case. Across these scenarios, unconditional generation reproduces geological realism, and conditional generation honors well data while producing geologically consistent between-well realizations, demonstrating practical utility for facies geomodelling applications.

1. **Introduction**

Geomodelling – constructing earth models consistent with geological knowledge, borehole interpretations, geophysical observations, and dynamic borehole observations – is essential for understanding geological processes, managing earth resources including underground water, $CO_2$ storage, minerals, and hydrocarbons, as well as making sustainable decisions. Three important criteria of geomodelling evaluation include realism (i.e., consistency with geological knowledge and spatial patterns), conditioning degree to the given data (e.g., borehole and geophysical data), and diversity (expressing variability and uncertainty) of the constructed geomodels. Creating geomodels with high realism, accurate conditioning performance, and sufficient diversity is crucial for informed decision-making in earth resource management.

Two typical geostatistics-based geomodelling methods are variogram-based (Oliver & Webster, 2014) and multi-point statistics (MPS; Strebelle 2002)-based approaches. They have been applied to reservoir geomodeling (e.g., Linde et al. 2015; Tahmasebi 2018) and geophysical inversion (Bosch et al., 2010), with their own advantages and disadvantages. In variogram-based methods, variograms representing spatial two-point relations are calculated to express geological patterns, and the constructed geomodels exhibit relatively weak geological realism. In contrast, MPS-based methods utilize multiple-point spatial relations to express geological patterns, thus leading to more realistic results as compared to variogram-based methods. However, the generated realism from MPS still needs improvement when modelling complex heterogeneous reservoirs (S. Song, Huang, et al., 2025).

Deep generative methods have proven capable of synthesizing realistic images (Dhariwal & Nichol, 2021; Ho et al., 2020; Karras et al., 2017) and videos (Voleti et al., 2024; T.-C. Wang et al., 2018). Commonly used generative methods (Bond-Taylor et al., 2022) include variational autoencoders (VAEs; Kingma and Welling 2013), flow-based models (Durk P Kingma & Dhariwal, 2018), generative adversarial networks (GANs; Goodfellow et al. 2014), and diffusion models (Ho et al., 2020).

Researchers have studied the application of GANs for geomodelling in various aspects in recent years (Chen et al., 2023; Cui et al., 2024; Laloy et al., 2018; Mosser et al., 2020; Nesvold and Mukerji, 2021; Song et al., 2021a, 2021b; Song et al., 2022b, 2022a; Zhang et al., 2019). After training a convolutional neural network model, called a generator, alongside another neural network called a discriminator, GANs can produce realistic reservoir geomodels by accepting random low-dimensional latent vectors as inputs. To further constrain these produced geomodels to given conditioning data, three types of methods were proposed: searching appropriate latent vectors that can result in conditional geomodels through optimization or sampling methods (e.g., Zhang et al. 2019; Nesvold & Mukerji, 2021), training an additional inference network to map another set of low-dimensional random latent vectors into the appropriate input latent vectors of the trained generator (Chan & Elsheikh, 2019), and training the generator such that it can directly take in any given conditioning data – in addition to the random latent vector – and produce realistic and conditional reservoir geomodels, as seen in GANSim (S. Song et al., 2021a; S. Song, Mukerji, & Hou, 2022). In recent years, GANSim has been further improved to combine with a deep learning-based flow surrogate model for conditioning to dynamic data (S. Song et al., 2023), to integrate a falsification scheme for modelling multiple geological scenarios, and to introduce a local discriminator design for better local realism around well facies data (S. Song, Mukerji, et al., 2025). GANSim has also been applied to the geomodelling of three-dimensional (3D) stationary-field karst cave reservoirs (Song et al., 2022b) and 3D non-stationary turbidite and delta reservoirs (Alqassab et al., 2024; S. Song, Mukerji, et al., 2025). However, training GANs may encounter difficulties in achieving stable convergence (Saxena & Cao, 2021) because it requires balancing the training of the discriminator and the generator in an adversarial way.

Recently, diffusion models have gained momentum in image generation tasks (Dhariwal & Nichol, 2021; Yang et al., 2024), video generation tasks (Xing et al., 2023), and audio generation tasks (Kong et al., 2020), among others. They have also been largely utilized in large artificial intelligence models for image or video generation from texts, such as Sora developed by OpenAI in 2014, DALL·E 2 (Ramesh et al., 2022), and Stable Diffusion 3 (Esser et al., 2024). Dhariwal and Nichol (2021) compared diffusion models with GANs, demonstrating that the former can produce realistic and diverse images with comparable or even better performance than the latter.

There are different versions of diffusion models, such as denoising diffusion probabilistic models (DDPMs; Ho et al. 2020), noise-conditioned score network (Y. Song & Ermon, 2019), and score-based generative modeling (Y. Song et al., 2020), among which DDPM is the most popular one at present. DDPM includes two processes: a fixed Markov chain forward process, gradually adding Gaussian noises to a training sample (e.g., a picture) through multiple steps to finally form a pure standard Gaussian noise, and a reverse process, gradually denoising a pure Gaussian noise through multiple steps to finally generate a data sample (e.g., a picture) which falls inside the distribution of the training datasets. During training, using the fixed forward process, a deep learning model (e.g., a convolutional neural network, CNN) is trained to predict the data distribution at a step closer to the training data (the earlier step), based on the data from the corresponding later step. Once trained, iteratively executing that deep learning model step by step would gradually transform standard Gaussian noise into a data sample within the training dataset distribution. DDPM performs all generation steps, which is computationally expensive during inference. Therefore, a sampling method called denoising diffusion implicit models (DDIMs; Song et al. 2022a) was proposed to speed up the inference process by skipping steps.

Researchers have applied diffusion models in geoscience problems, including geomodelling (Di Federico & Durlofsky, 2024, 2025; Lee et al., 2025), seismic velocity synthesis (F. Wang et al., 2024), and digital rock reconstruction (Luo et al., 2024). For example, Lee et al. (2025) applied a latent diffusion model proposed by Rombach et al. (2022), to simulate two-dimensional (2D) vertical facies distribution sections for a shoreface reservoir, conditioned to well interpretations (as strings of discrete facies codes along wells). The "classifier-free guidance" proposed by Ho and Salimans (2022), as well as an additional "preservation" loss comparing the input conditioning well facies data and the produced facies model, are used to train the diffusion model to achieve conditioning effect.

The preservation loss proved to be one of the key factors for conditioning. The design of the preservation loss is nearly identical to the conditioning loss for well facies data in GANSim (S. Song et al., 2021a). However, the geological patterns in the training dataset of Lee et al. (2025) are relatively monotonous, and tuning the weights of different loss terms is required. Di Federico and Durlofsky (2024) developed diffusion-based latent parameterizations for geomodels and imposed conditioning through flow-based history matching; their subsequent work extends this framework to 3D latent diffusion (Di Federico & Durlofsky, 2025). While effective for calibrating models to dynamic (flow) responses, this assimilation-based conditioning differs from the geomodelling setting emphasized here, where the goal is to directly generate facies realizations that honor sparse, hard geological constraints (well facies) as exact local information.

In this paper, inspired by image inpainting and outpainting techniques proposed by Saharia et al. (2022), we propose a mask-based conditioning formulation for well facies interpretations that treats conditioning as a masked generation problem within the diffusion framework. Rather than introducing an explicit preservation loss, the conditioning is incorporated through the masking mechanism during training and sampling. We evaluate this approach on three reservoir types with diverse geological patterns, and the results suggest that the method provides a practical and generally applicable way to perform well-based conditioning in diffusion-model facies geomodelling. The proposed methodology is outlined in Section 2. In Section 3, we present the training datasets for the three types of reservoirs. In Section 4, the unconditional and conditional generation results obtained with the proposed method are presented and analyzed. Finally, Sections 5 and 6 provide the discussions and conclusions, respectively.

## 2. Geomodelling method based on diffusion models

In this section, we first briefly introduce the work steps of DDPM and DDIM. Then, we show how the two methods are used for unconditional geomodelling (i.e., DDPM for training and DDPM/DDIM for inference). Next, a mask-based conditioning method based on diffusion models is proposed for conditional geomodelling on well facies data.

### 2.1 Brief introduction of DDPM and DDIM

According to Ho et al. (2020), the work steps of DDPM are as shown in Table 1. During the training stage, first, $x_0$, either a two-dimensional image or a three-dimensional cube, is randomly selected from the training dataset $q(x_0)$. Then, choose a random time step $t \in \{1, 2, 3, ..., T\}$ among all pre-defined time steps involved in the forward or inverse process of DDPM. Next, sample a Gaussian noise $\epsilon$ from $N(\mathbf{0}, \mathbf{I})$, where $\mathbf{0}$ and $\mathbf{I}$ are zero and identity matrices with the same dimension as a training sample. Fourth, calculate the data of the time step $t$, $x_t$, from $x_0$ using

$$x_t = \sqrt{\bar{a}_t} x_0 + \sqrt{1 - \bar{a}_t} \epsilon, \tag{1}$$

where $\bar{a}_t = \prod_{i=1}^{i=t} \alpha_i$, $\alpha_t = 1 - \beta_t$, and $\beta_t$ is a predefined parameter controlling the noise addition schedule for each time step of the forward process. Then, take $x_t$ and $t$ as inputs into a neural network $\epsilon_\theta$ ($\theta$ are the trainable parameters) and train the neural network with a loss

$$L = E_{x_0 \sim q(x_0), \epsilon \sim N(\mathbf{0}, \mathbf{I}), t \sim [1, T]}[||\epsilon - \epsilon_\theta(x_t, t)||^2], \tag{2}$$

where $E$ is the expectation operation. This neural network $\epsilon_\theta$ is trained to predict the added noise $\epsilon$ and is often referred to as the denoising network, or simply the noise predictor. Finally, repeat the above work steps to train the denoising network $\epsilon_\theta$ until the loss function converges.

After training, the noise predictor $\epsilon_\theta$ can be used to sample the intermediate data at a previous time step ($x_{t-1}$) given the data of the later time step ($x_t$) using

$$x_{t-1} = \frac{1}{\sqrt{\bar{a}_t}} \left( x_t - \frac{1 - \alpha_t}{\sqrt{1 - \bar{a}_t}} \epsilon_\theta(x_t, t) \right) + \sigma_t z, \tag{3}$$

where $\sigma_t$ is assumed to be equal to $\beta_t \frac{1 - \bar{a}_{t-1}}{1 - \bar{a}_t}$, and $z \sim N(\mathbf{0}, \mathbf{I})$ has the same dimension as the training data. Thus, once a Gaussian noise is sampled at the final time step, i.e., $x_T \sim N(\mathbf{0}, \mathbf{I})$, then the corresponding $x_0$ can be generated by iterative execution of Equation (3) in reverse order from the final to the initial time step. The generated $x_0$ is expected to fall within the training data distribution (Ho et al., 2020).

**Table 1.** The pseudo-code of the training and inference process of DDPM (adapted from Ho et al. (2020))

| Algorithm 1 Training | Algorithm 2 Sampling |
|---|---|
| 1: repeat | 1: $x_T \sim N(0, I)$ |
| 2: $x_0 \sim q(x_0)$ | 2: **for** t = T, ..., 1 **do** |
| 3: t ~ Uniform({1, ..., T}) | 3:    if t > 1, $z \sim N(0, I)$ |
| 4: $\epsilon \sim N(0, I)$ | 4:    else, $z = 0$ |
| 5: $x_t = \sqrt{\bar{a}_t} x_0 + \sqrt{1-\bar{a}_t} \epsilon$ | 5:    $x_{t-1} = \frac{1}{\sqrt{\bar{a}_t}} \left( x_t - \frac{1-\alpha_t}{\sqrt{1-\bar{a}_t}} \epsilon_\theta(x_t, t) \right) + \sigma_t z$ |
| 6: Train the neural network $\epsilon_\theta$ with loss $\|\epsilon - \epsilon_\theta(x_t, t)\|^2$ | 6: end for |
| 7: **until** converged | 7: return $x_0$ |

Iteratively applying the well-trained noise predictor ($\epsilon_\theta$) across all time steps is computationally expensive during inference. Therefore, the denoising diffusion implicit model (DDIM; Song et al. 2022a) was proposed for inference with significantly fewer time steps. The derivation of DDIM removes the Markov-chain assumption required in DDPM while maintaining the same loss function and training steps as DDPM. The noise predictor $\epsilon_\theta$ trained in DDPM can be used for inference in DDIM. The deterministic sampling equation used for inference in DDIM is

$$x_{t-k} = \sqrt{\bar{a}_{t-k}} \frac{x_t - \sqrt{1-\bar{a}_t} \epsilon_\theta(x_t, t)}{\sqrt{\bar{a}_t}} + \sqrt{1-\bar{a}_{t-k}} \epsilon_\theta(x_t, t), \tag{4}$$

where $k$ is the number of skipped time steps. More details of the theories about DDPM and DDIM can be found in Ho et al. (2020) and Song et al. (2022a).

2.2 Unconditional geomodelling with DDPM and DDIM

We apply DDPM and DDIM for unconditional geomodelling of three types of reservoirs: two in 2D and one in 3D. The 2D reservoirs are discretized into 64×64 cells, containing three and four facies categories, respectively. The 3D reservoir is partitioned into 48×48×32 cells and includes four facies types. The following Data Section provides further details on the reservoir and training datasets. Figure 1 shows a schematic diagram of the forward (from geomodel to noise) and the reverse (from noise to geomodel) processes of DDPM involved in unconditional geomodelling.

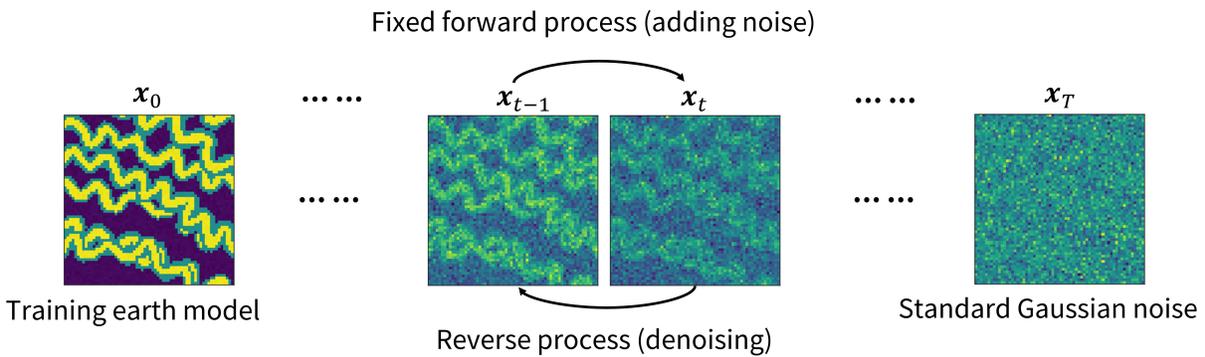

**Figure 1.** A schematic diagram of the forward (from geomodel to noise) and the reverse (from noise to geomodel) processes of DDPM involved in unconditional geomodelling.

The architecture of the denoising network $\epsilon_\theta$ used for unconditional geomodelling in this paper essentially follows a U-Net design. Figure 2 illustrates the architecture of $\epsilon_\theta$ for the 2D reservoir cases as an example. It is composed of three encoding blocks (denoted as Encoding Block), one middle bottleneck block (denoted as Mid Block), and three decoding blocks (denoted as Decoding Block). There is a downsample layer (denoted as Down sample) between every two consecutive encoding blocks, and an upsample layer (denoted as Up sample) between every two consecutive decoding blocks. A residual block (denoted as Res Block) is designed to closely follow the last decoding block. Skip connections are established between the encoding and decoding blocks at each resolution level. The input time step value *t* is initially processed through a time multi-layer perceptron block (denoted as time MLP). The output of the perceptron block is processed by time-embedding blocks (denoted Time Embedding), and the resulting time-step information is subsequently fed into different blocks of the U-Net. The convolutional layers with a 1x1 kernel size, denoted 1x1 Conv, are inserted to process input and output information. The input and output sizes of the U-Net, as well as the sizes of intermediate feature maps after key neural blocks, are shown in Figure 2. Further illustrations of the design of each block are illustrated in Appendix A.

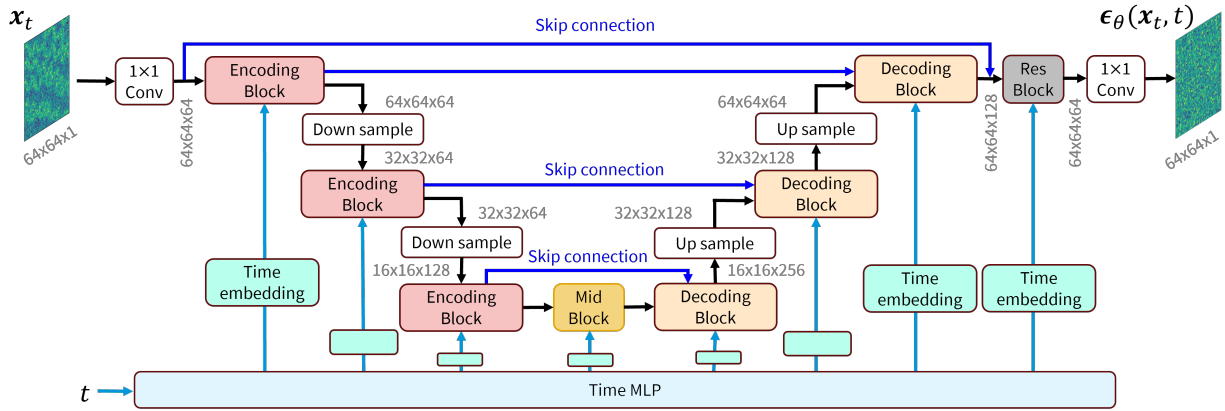

**Figure 2.** The architecture of the noise predictor $\epsilon_\theta$ employed for unconditional geomodelling of the two 2D cases in this paper. Further details of each neural block are presented in Appendix A.

Finally, with the U-net design of the noise predictor $\epsilon_\theta$ and following the worksteps outlined in Table 1, $\epsilon_\theta$ can be trained and iteratively sampled to produce unconditional geomodels with DDPM. For faster inference, DDIM can also be employed.

2.3 Mask-based conditioning method for well facies data

To condition on sparse well facies data, a mask-based method is proposed. Compared to the previous unconditional geomodelling method, there are two major differences. First, the sparse conditioning well facies information is incorporated into the denoising network $\epsilon_\theta$ through indicator-based encoding. As illustrated in Figure 3, the well facies data ($W$) are decomposed into multiple facies indicator maps ($W_{\text{ind}}$)—one for each facies category—and one well location indicator map ($m$). The well location indicator map takes the value of 1 at well locations and 0 elsewhere. Each facies indicator has a value of 1 if the corresponding facies is present at the well location, and 0 otherwise (i.e. "one-hot encoding"). All the intermediate data at different time steps ($x_t$) are masked by the original sparse well facies codes ($W$) at well locations with

$$x_t^m = (1-m)\otimes x_t + m\otimes W, \qquad (5)$$

where $\otimes$ is the element-wise multiplication operator. The masked intermediate data $x_t^m$, along with $W_{\text{ind}}$ and $m$, is further concatenated and fed into the noise predictor network $\epsilon_\theta$ during training. Such an operation, as in Equation (5), is referred to as well facies masking in this paper.

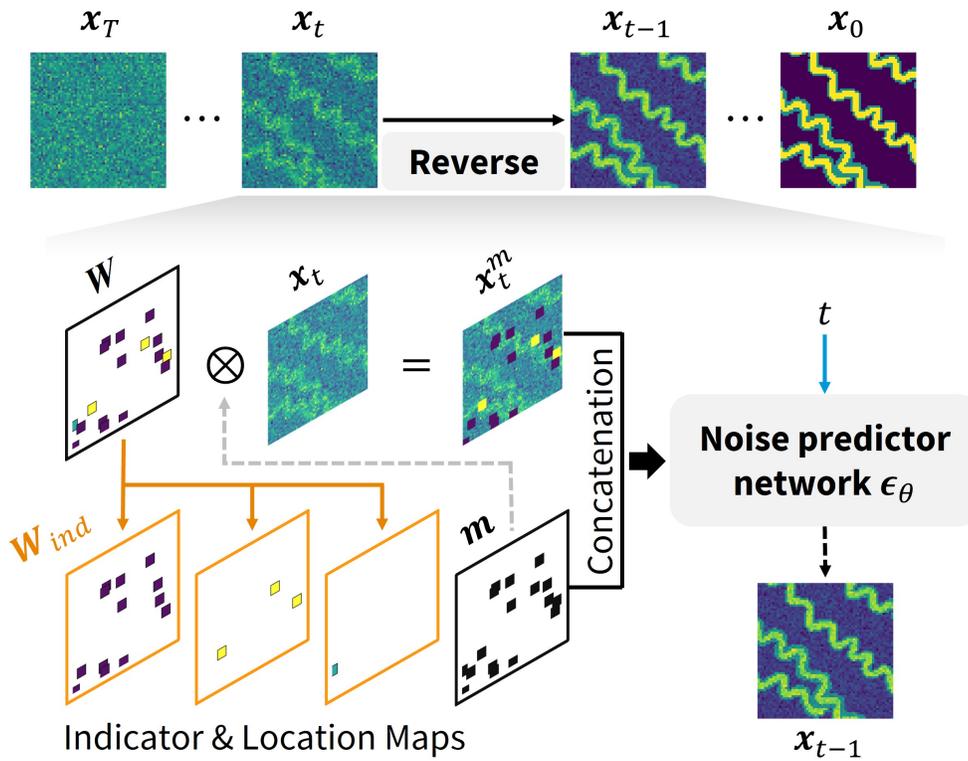

**Figure 3.** The conditional facies simulation using diffusion models incorporating conditional

indicators and masked intermediate data as inputs. The facies points on the bottom facies models are enlarged from 1x1 to 5x5 pixels for visualization purposes.

Second, the loss function is modified to only denoise the between-well region of $x_t^m$:

$$L_W = E_{x_t^m, m, W_{\text{ind}}, \epsilon, t} \left[ (1 - m) \otimes ||\epsilon - \epsilon_\theta(x_t^m, m, W_{\text{ind}}, t)||^2 \right]. \tag{6}$$

The loss refers to the difference between the predicted and the true noise only for between-well regions. Unlike the method of additional well facies mismatch loss (i.e., the "preservation" loss) for conditioning to well facies data in Lee et al. (2025), this method eliminates the need to balance the weights of different loss terms. This leads to a more straightforward approach without compromising the realism and conditioning effect of the generated geomodels.

During inference, the conditioning well facies information is injected into the trained noise predictor at each time step in the same way as during the training. Following the worksteps of DDPM or DDIM, the final facies realizations that align with expected geological patterns and the input well facies data can be generated.

## 3. Data

Three datasets of channel reservoir models are utilized to evaluate the geomodelling performance of the proposed diffusion model methods, both in unconditional and conditional scenarios. The first dataset is constructed by Song et al. (2021a) using object-based modeling in the Petrel platform, and is referred to as the channel reservoir data. Figure 4a shows several facies model examples with diverse channel orientation, sinuosity, and facies proportion across different models. There are three classes of facies: channel sand, channel bank, and inter-channel mud. Each facies model consists of multiple channels with similar geometric features, such as channel orientation, width, and sinuosity. The dimension of each facies model is 64 x 64 cells, with each cell representing an area of 50 x 50 m. The entire facies model dataset is split into a training set of 32,640 samples and a test set of 3,000 samples. We randomly sampled 3 to 32 well locations, each with a 1x1 cell, from each facies model to serve as conditioning well facies data. Some examples of facies models and the corresponding well facies data are presented in Figures 4a and 4b.

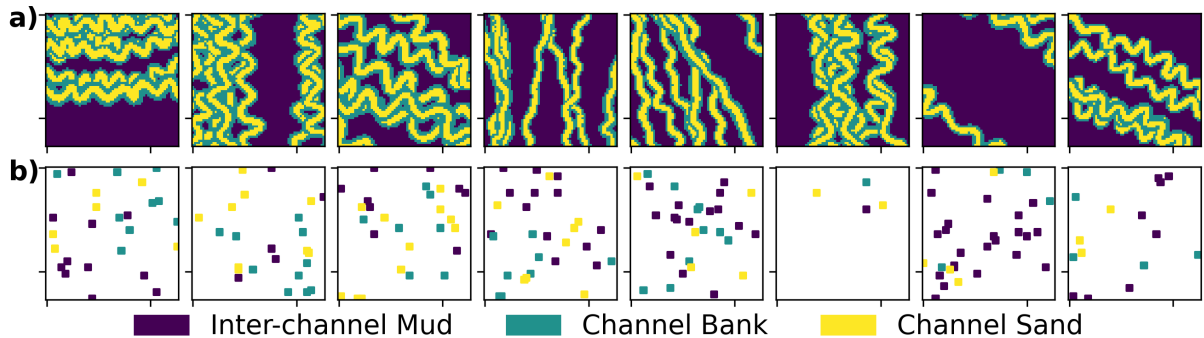

**Figure 4.** Some facies model examples of the channel reservoir data. a) presents facies models, while b) illustrates the corresponding well facies data. There are three types of facies: channel sand (yellow), channel bank (teal), and inter-channel mud (purple). The well facies points in b) are enlarged from 1x1 pixels to 5x5 cells in this figure for better visualization.

The second facies model dataset used in this paper was created by Hu et al. (2024) and includes four facies types: abandoned channel, lateral accretion sand, mud drape, and floodplain. It is referred to as the point bar reservoir data. Illustrative examples of facies models and the corresponding well facies data are shown in Figures 5a and 5b, respectively. Each facies model consists of 64 x 64 cells, where each cell corresponds to 10 x 10 m. These data pairs are divided into a training data set including 10,000 data samples and a test data set including 1700 data samples. We randomly selected 4 to 31 well locations, each with a 1x1 cell, from each facies model as the conditioning well facies data.

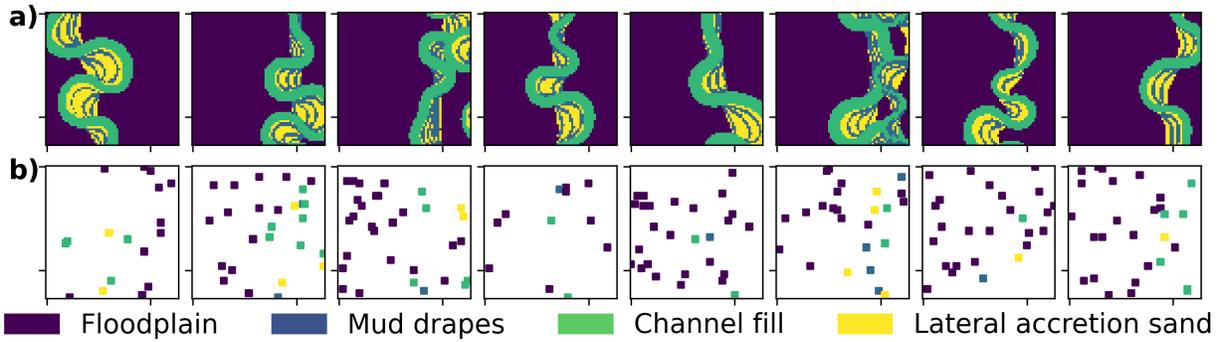

**Figure 5.** Some examples of the point bar reservoir dataset. a) presents facies models, while b) illustrates the corresponding well facies data. There are four types of facies: channel fill (sea green), lateral accretion sand (yellow), mud drapes (dark blue), and floodplain (purple). The well facies points in b) are enlarged from 1x1 pixels to 5x5 pixels in this figure for better visualization.

The third dataset used in this study originates from 200 3D meandering channel models generated by Hu et al. (2024). Each model consists of 200×200×160 cells along the x, y, and z directions with a resolution of 10×10×0.1 m. They have the same facies classes as the previous point bar dataset. We randomly select 190 out of the 200 models to construct a training dataset, while reserving the remaining 10 for testing. By randomly cropping along abandoned channel facies, we get 12,000 facies model patches, each containing 48x48x32 cells, in the training dataset and 500 patches in the test dataset. From each facies-model patch, we randomly select between 3 and 31 well locations to serve as conditioning data. Each well is represented as a vertical column of size 1x1x32 containing facies codes. Some examples of the facies model patches and the corresponding well facies data are displayed in Figure 6, where the floodplain facies are omitted in the first two rows for improved visualization.

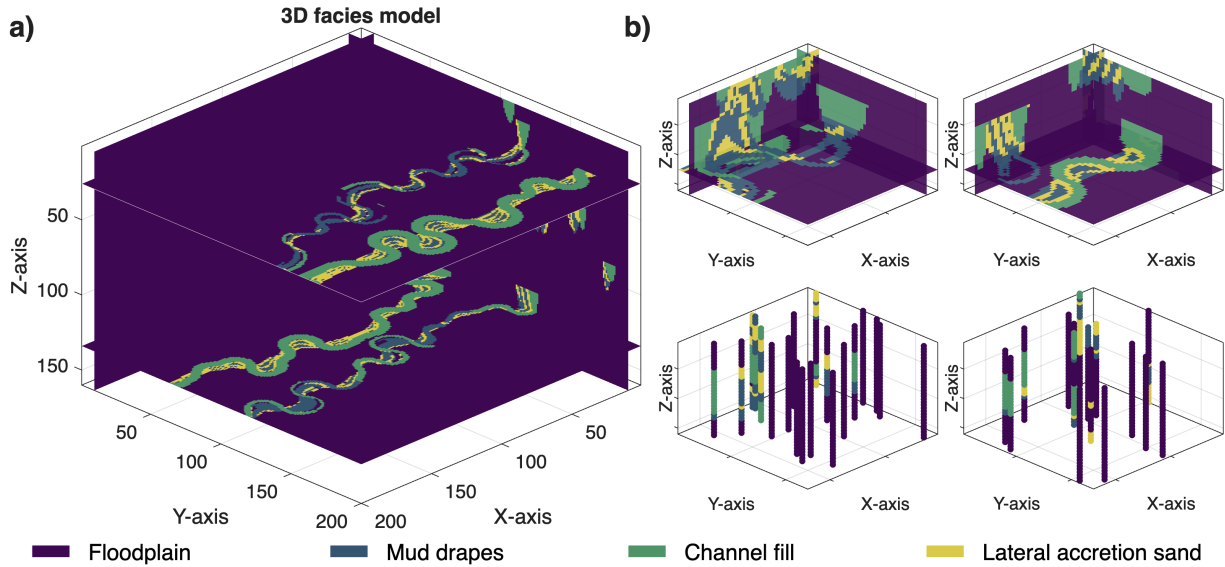

**Figure 6.** Examples from the 3D point-bar dataset. Panel a) shows an original 3D facies model, and panel b) shows four extracted training samples (top row) along with the corresponding well facies data (bottom row).

## 4. Results and analyses

### 4.1 Unconditional geomodelling

#### 4.1.1 Channel reservoir case

The architecture of the noise predictor network is presented in Figure 2. Following the training procedure outlined in Table 1, we train the noise predictor network by using the Adam optimizer , with a batch size of 256, and 1500 denoising iteration steps. The denoising-step setting is selected based on an ablation study (Section 5), as it significantly impacts facies generation performance. The entire training process, spanning 300 epochs and approximately 38,000 mini-batch iterations, takes about 8 hours to complete in a single Nvidia A100 GPU with 80G of memory. The original and smoothed loss curves presented in Figure 7 indicate that the training converges well. Based on this we select the model with the lowest epoch loss for inference. After training, the noise predictor network is expected to capture the geological patterns embedded in the training samples and can be applied to generate realistic geomodels.

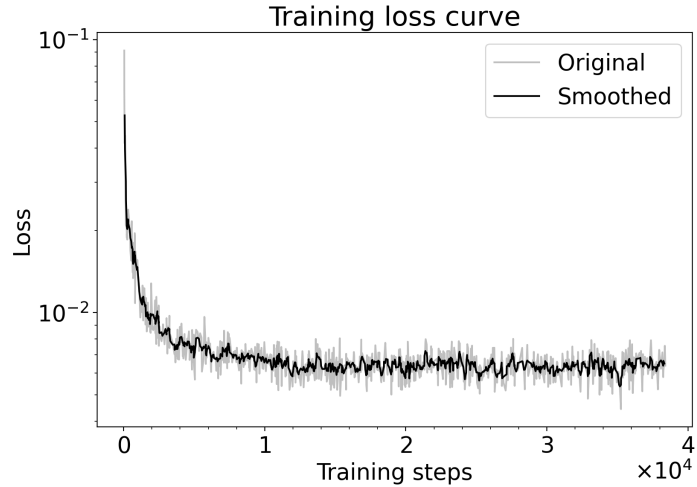

**Figure 7.** The evolution of both the original and smoothed loss values over iteration steps during the training of the noise predictor network.

Based on the trained noise predictor, Equation (3) (if using DDPM) and Equation (4) (if using DDIM) are iteratively performed from the last time step to the initial time step to generate a geomodel from a pure Gaussian noise map. Figure 8 shows the DDPM inference results for the time steps 1, 700, 1000, 1400, and 1500 for three random Gaussian noise maps. When using DDIM for inference, we set the skipped time step number (k in Equation (4)) as 30, so a total of 50 iterations of executions of Equation (4) are required. Figure 9 presents DDIM results after different execution iterations for three random Gaussian inputs. From the three randomly selected generation processes in Figures 8 and 9, we can see that DDPM and DDIM both can progressively generate realistic facies models from pure Gaussian noise maps. The average inference time for each geomodel is about 0.8 seconds for DDPM and 0.029 seconds for DDIM, respectively, with DDIM giving an order of magnitude speed-up during inference . Further details regarding the setting of training time steps are provided in the discussion section.

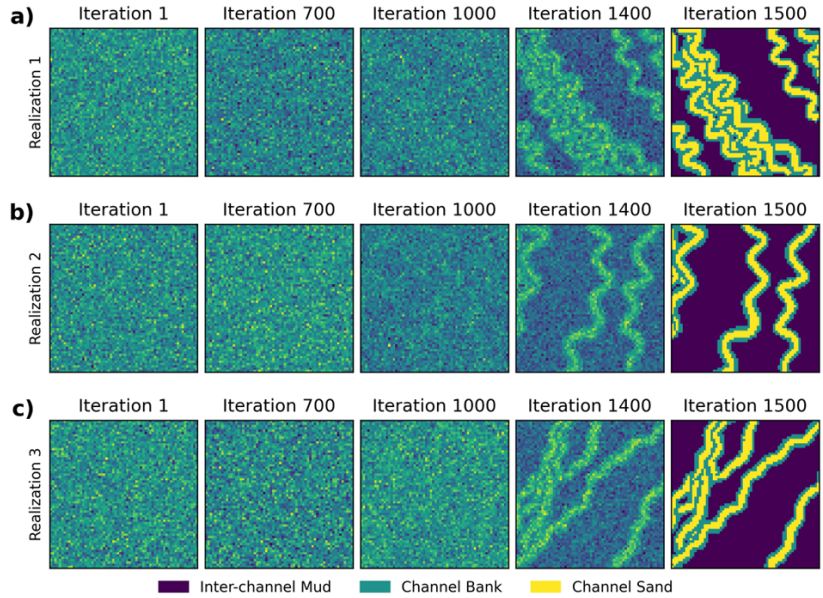

**Figure 8.** The sampled outputs of the iterations 1, 700, 1000, 1400, and the final 1500 with three different Gaussian noise maps as the initial inputs following the DDPM inference procedure.

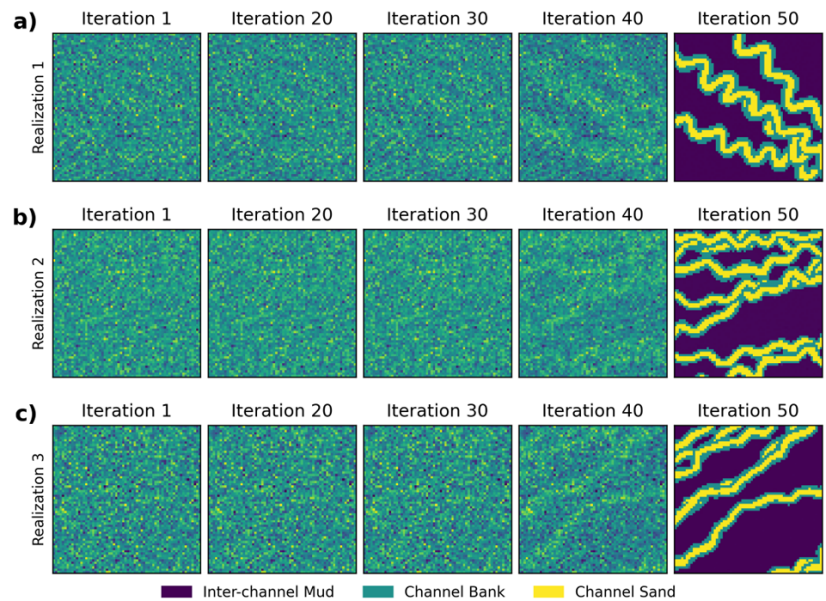

**Figure 9.** The sampled outputs of iterations 1, 20, 30, 40, and the final 50 with three different Gaussian noise maps as the initial inputs following the DDIM inference procedure. In each iteration, 30 time steps are skipped.

To further investigate generation performance quantitatively, we simulate two sets of 12,000 facies models using both DDPM and DDIM, respectively, and compare their distributions with that of the 3000 test set facies models. Since each model is a high-dimensional object, the comparison in done by projecting the models to a 2D space based on multidimensional scaling (MDS; Borg and Groenen, 2005) using a multi-scale sliced-Wasserstein distance (MS-SWD; Karras et al., 2017). It includes two steps: first, we divide the test and the generated facies models into multiple groups each containing 100 facies models; then, calculate the MS-SWD between every two facies model groups of the test set and the generated realizations of either DDPM or DDIM, and finally, project these facies model groups onto a two-dimensional MDS space based on the calculated MS-SWD matrix with each group represented by a point. More details of this combined method are described in Song et al. (2021a). The projected facies model groups are visualized using scatter plots and density plots in Figure 10 for DDPM and Figure 11 for DDIM. The strong consistency between the generated and test facies model distributions demonstrates that both DDPM and DDIM can effectively capture the underlying spatial geological patterns of the training geological models.

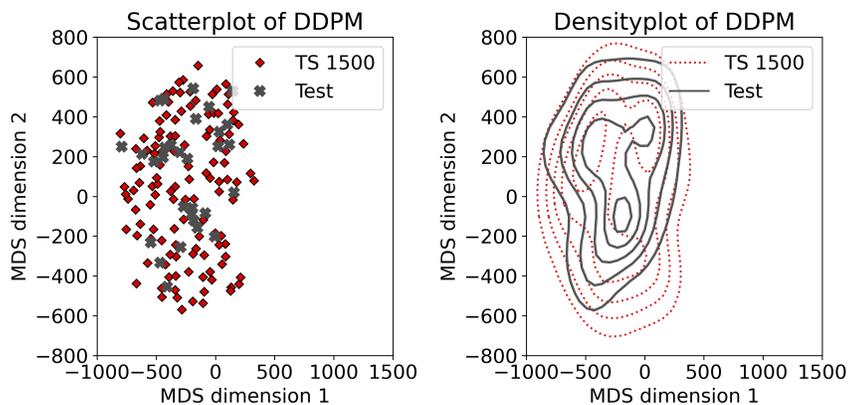

**Figure 10.** Multidimensional scaling plot of the test set and realizations generated using DDPM with 1500 time steps. 'TS' (timesteps) in the legends denotes the total number of timesteps for diffusion model training. The left panel shows a scatter plot, while the right panel presents a density plot of the MDS projections.

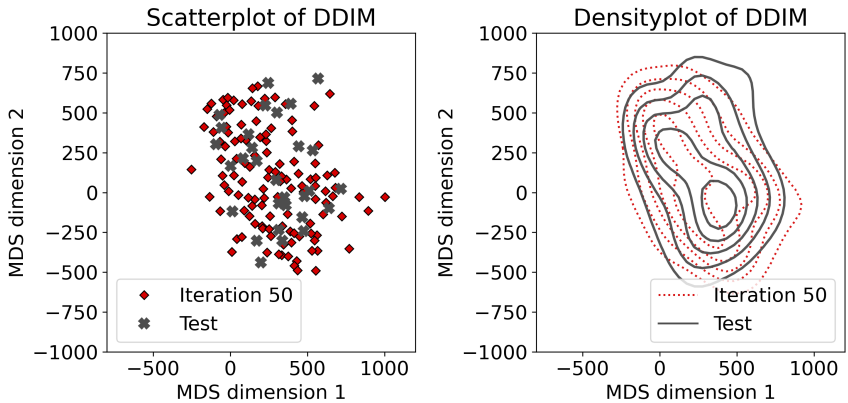

**Figure 11.** Multidimensional scaling plot of the test set and realizations generated using DDIM with 50 sampling steps.

We further compare the generated results with the test dataset using facies proportion distributions, variograms, and channel geometric features. Figure 12 compares the facies proportion histograms of the three facies types (inter-channel mud, channel bank, and channel sand facies) of the test dataset, including 3000 facies models, as well as the generated realizations from DDPM and DDIM, each including 3,000 samples. The histograms of the generated facies models match those of the test dataset very well.

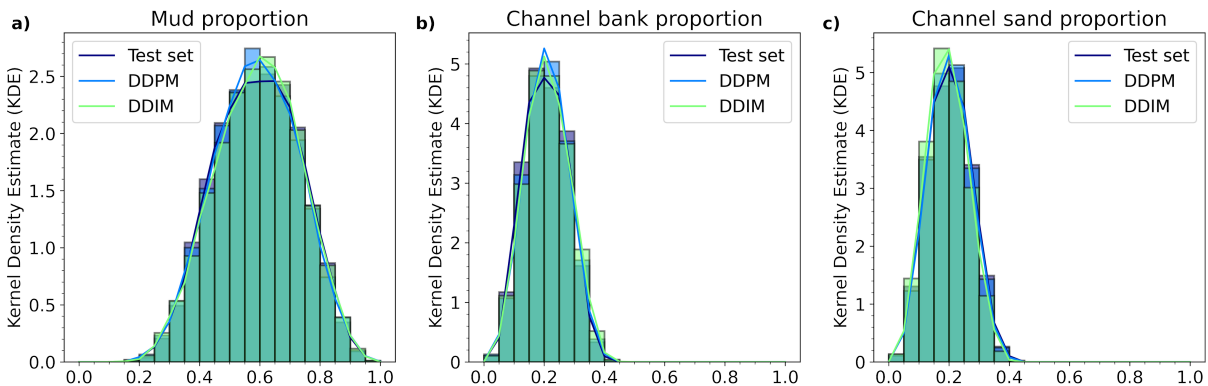

**Figure 12.** The class distribution comparison of three types of facies between the test set and the generated realizations generated by DDPM and DDIM. a)-c) compare the class distributions of the inter-channel mud, channel bank, and channel sand, respectively.

We also randomly selected 400 facies models from the test set, DDPM-generated and DDIM-generated realizations, respectively, and computed the variograms of each selected facies model along four directions – 0°, 45°, 90°, and 135° in a clockwise direction from the north. Figure 13 shows the distributions of these calculated variograms. The solid and dashed curves are the mean of the 400 variograms of the 400 test and generated facies models, while the shaded regions show the range between the 10th and 90th percentiles of the 400 variograms. We can see that the variograms of the generated facies models match very well with those of the test ones.

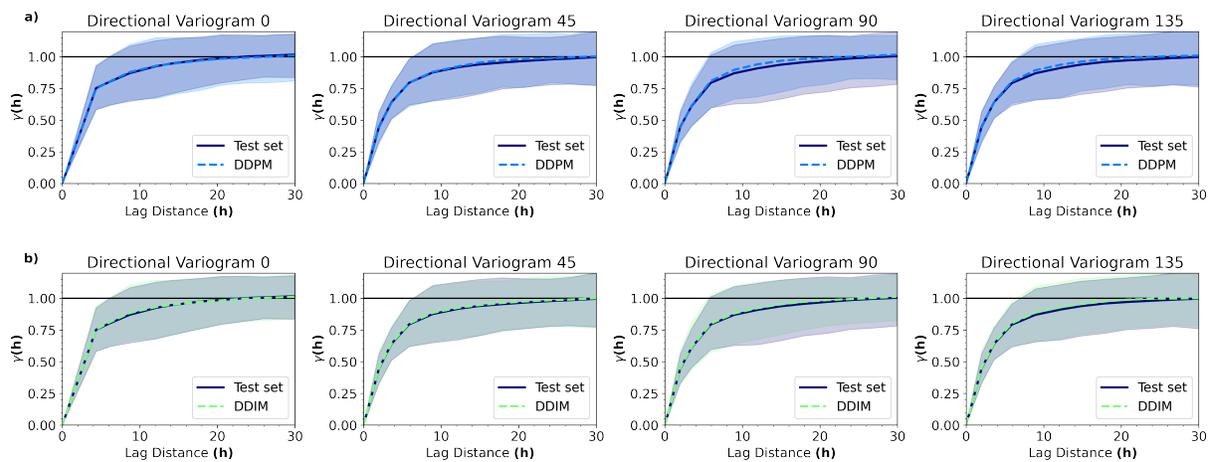

**Figure 13.** The variogram comparison along four directions between 400 random geomodels from the test set and 400 random realizations generated from a) DDPM and b) DDIM. The solid or dashed lines represent the mean value of 400 variograms, while the shaded area surrounding the mean value curve signifies the range between the 10th and 90th percentiles of 400 variograms.

We further compare the distributions of the channel length, sinuosity, and width between the generated and the test facies models. For each facies model, we select only the longest channel to measure these geometric features. Sinuosity is calculated as the actual channel length divided by the straight-line distance between the two endpoints of the channel (Fuller et al., 2013). The geometric feature distributions (as cumulative density functions) of 3000 test facies models and 3000 realizations from DDPM and DDIM are compared in Figure 14. The geometric feature distributions of the DDPM-generated facies models are closer to those of the test facies models than the DDIM-generated results. Nonetheless, both DDPM and DDIM exhibit overall distributions that are largely consistent with the test dataset.

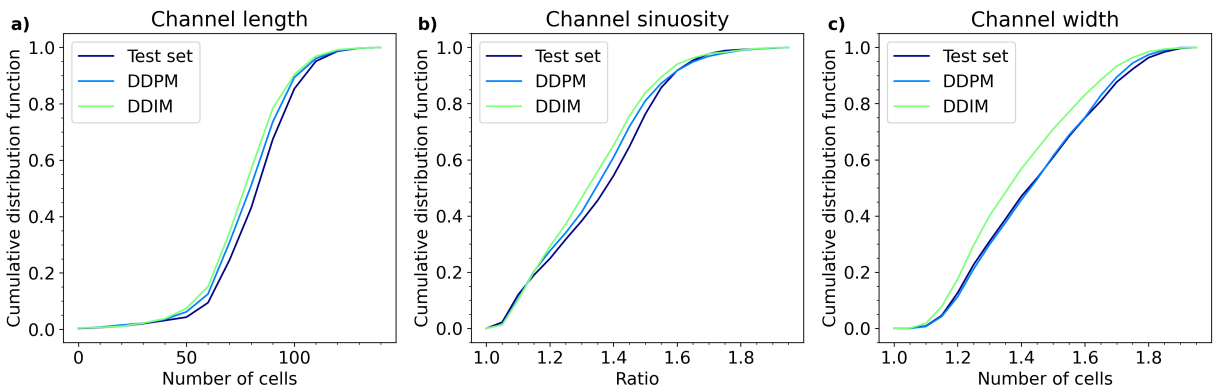

**Figure 14.** The geometric features comparison between the test set and the generated realizations from DDPM and DDIM. a)-c) display the distribution of channel length, channel sinuosity, and channel width.

Based on the above evaluation of the generated geomodels across various metrics, we find that diffusion models, both DDPM and DDIM, are capable of producing realistic and diverse geomodels. Notably, DDIM offers comparable geomodel quality to DDPM but with largely reduced inference time, making it a more efficient option for generating geomodels.

4.1.2 Point bar reservoir case

Using the same hyperparameter settings, neural network architecture, and training procedures as in the previous channel-reservoir scenario, we train the network for the point-bar case for 11 hours on an Nvidia A100 GPU. Then, the trained noise predictor network can be used for unconditional geomodelling. We generate 6,000 facies models using both DDPM (1500 denoising steps) and DDIM (50 sampling steps, skipping 30 steps between successive steps). Figures 15a and 15b present random realizations generated by DDPM and DDIM, respectively. Figures 15c and 15d present scatter and density plots, respectively, comparing the overall data distributions of the generated facies models and the test data using the combined MS-SWD and MDS method. The results demonstrate that both DDPM and DDIM are effective in generating geomodels that closely resemble the test facies models in terms of realism and diversity. Regarding data distribution in Figures 15c-15d, DDPM achieves relatively better performance than DDIM, albeit with increased computational time, possibly due to the larger number of inference iterations.

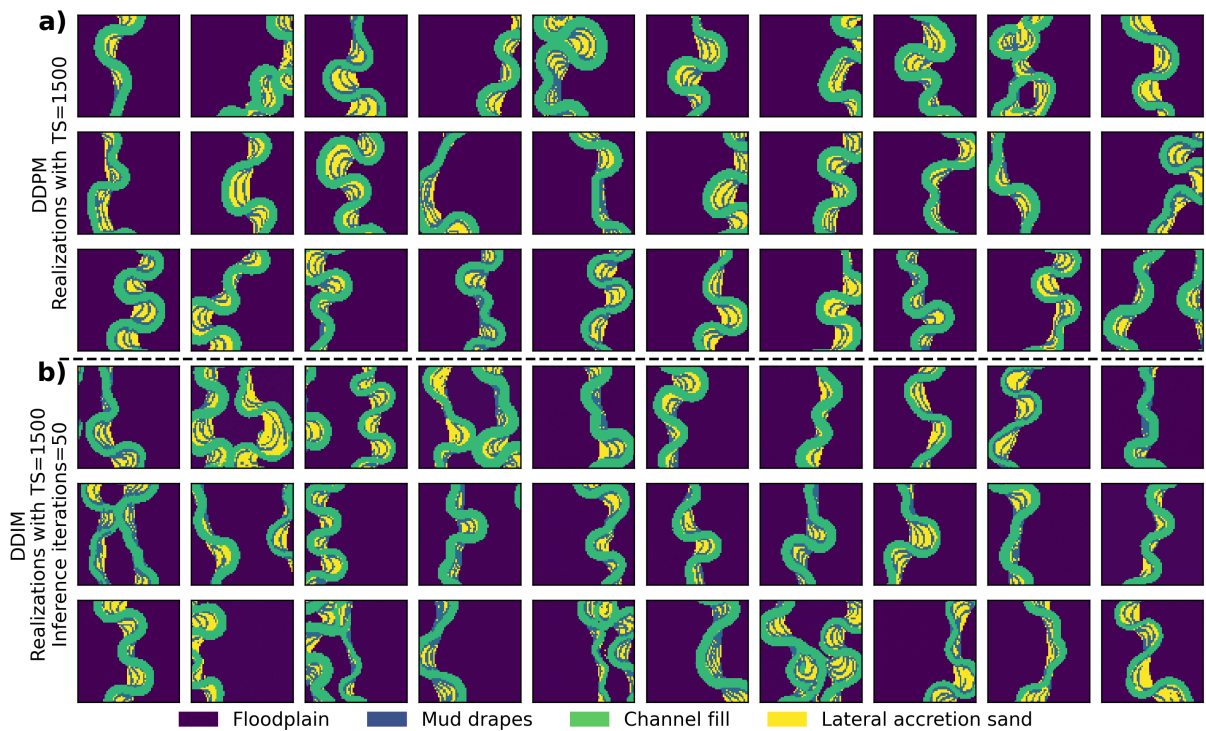

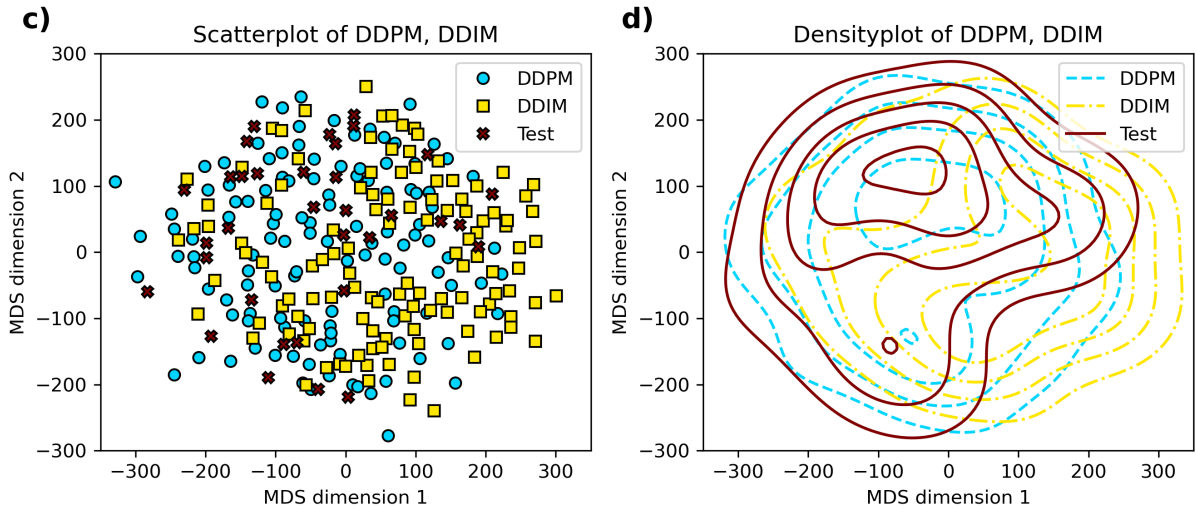

**Figure 15.** Unconditional realizations of the point bar reservoir geomodels generated using a) DDPM with 1500 time steps and b) DDIM with 50 sampling steps. The c) scatter plot and d) density plot illustrate data comparisons between the test samples and the generated ones with DDPM and DDIM.

4.1.3 Three-dimensional point bar reservoir case

For the 3D case, we adopt a U-Net architecture for the denoising generative model, following the same overall design as in the 2D experiments. As illustrated in Figures 2 and A1, the 3D U-Net retains the same building blocks and connectivity (e.g., residual blocks, attention modules, skip connections, and time conditioning), with all 2D convolutional layers replaced by their 3D counterparts to accommodate volumetric inputs. Other settings and procedures remain the same as the 2D cases, and the batch size is reduced to 32 due to GPU memory constraints. The denoising network is trained for 152 epochs, requiring approximately 20 hours on a single Nvidia A100 GPU. Following training, DDPM (with 1500 time steps) and DDIM (50 skip steps with each skip 30 time steps) are employed to generate the unconditional facies models, which are shown in Figure 16.

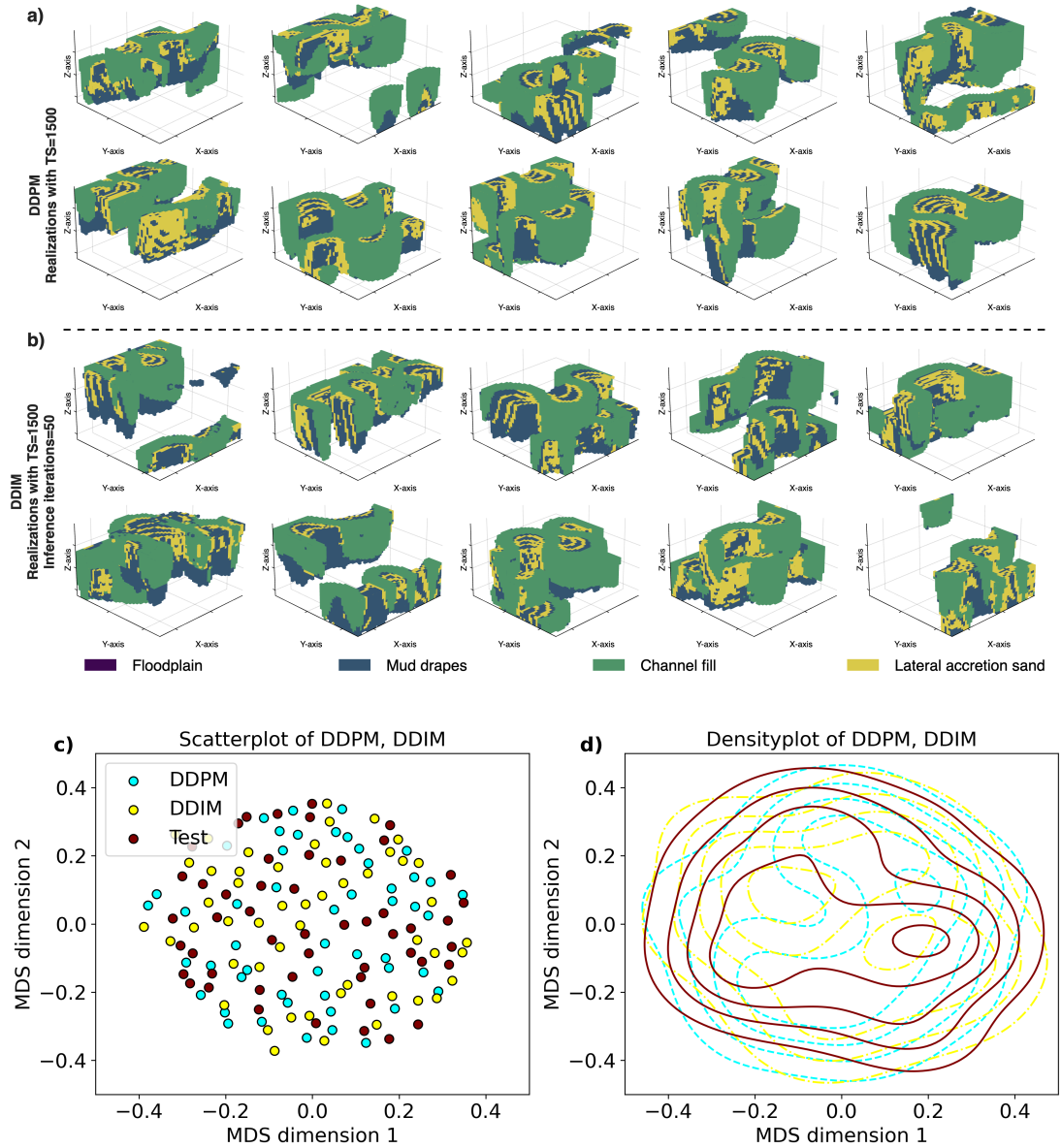

**Figure 16.** Unconditional realizations of 3D point bar reservoir case, generated using a) DDPM with 1500 time steps and b) DDIM with 50 sampling steps. The c) scatter plot and d) density plot illustrate data comparisons between test geomodels and the generated ones produced from DDPM and DDIM.

The data distribution comparisons in Figure 16 c-d are conducted using 50 test geomodel patches, 50 DDPM-generated ones, and 50 DDIM-generated samples. Whether compared based on visual inspection or distributions in MDS plots, the generated realizations present consistent spatial features to the test samples, indicating the effectiveness of the diffusion model-based geomodelling method for 3D reservoir cases. The inference time for DDPM (1500 denoising steps) is 3.5 minutes when sampling with a batch size of 8 running on an Nvidia A100 GPU, whereas DDIM (50 sampling steps) requires only 6.8 seconds under the same conditions.

4.2 Conditional facies modeling using the mask-based conditioning method

The proposed mask-based well facies conditioning method is validated using the two 2D datasets and the one 3D dataset. For the 2D cases, we use the architecture illustrated in Figure 2, combined with the mask-based conditioning approach for well facies data described in Figure 3, to train the denoising network. The number of denoising steps is set to 1500, and the batch size is set to 64. The training of the denoising network takes approximately 24 hours on a single Nvidia A100 GPU with 80 GB of memory in both 2D cases. After training, we utilize the trained denoising network for conditional geomodelling in the way of DDIM. In the 3D case, the same well facies mask-based method is integrated into the architecture of the unconditional scenario. We train the denoising network for approximately 32 hours on four Nvidia A100 GPUs. After training, both DDPM and DDIM can be used for conditional geomodelling using the trained denoising network. For conciseness, we provide the DDPM conditional results for three scenarios in the Supporting Information (Figures S1–S3), and we focus on DDIM in the main text.

4.2.1 Channel reservoir case

To explore the diversity of the generated realizations and assess the conditioning capability of the proposed method, we randomly select eight test facies models and use their extracted well facies data to generate 100 facies model realizations for each of the test facies models by varying the input Gaussian noise map. The results are illustrated in Figure 17. Each row of Figure 17 includes the test facies model, the corresponding input well facies data, six randomly generated geomodels, and the frequency maps of three facies types calculated from the 100 generated realizations. We can see from Figure 17 that, no matter how many conditioning wells there are, all generated facies models exhibit geological features comparable to those of the training dataset and condition to the input well facies data accurately. When constrained by fewer wells, such as the top three examples of Figure 17, the facies frequency maps globally exhibit a blurry and smooth feature except around well locations. This indicates that the generated facies model realizations possess sufficient diversity—e.g., all directions of generated channels. As the conditioning well number increases, as in the bottom five examples of Figure 17, well facies data interact with one another, thereby reducing the overall diversity of the generated facies models.

Upon visual inspection of the 800 realizations generated for the eight geomodeling cases in Figure 17, we observe that a small proportion (approximately 0.25%) of the generated facies models exhibit artifacts in the form of channel discontinuities, deviating from the continuous channel patterns present in the training dataset. Figure 18 illustrates two such examples, where the white arrows indicate the locations of channel disconnenctions in the generated realizations.

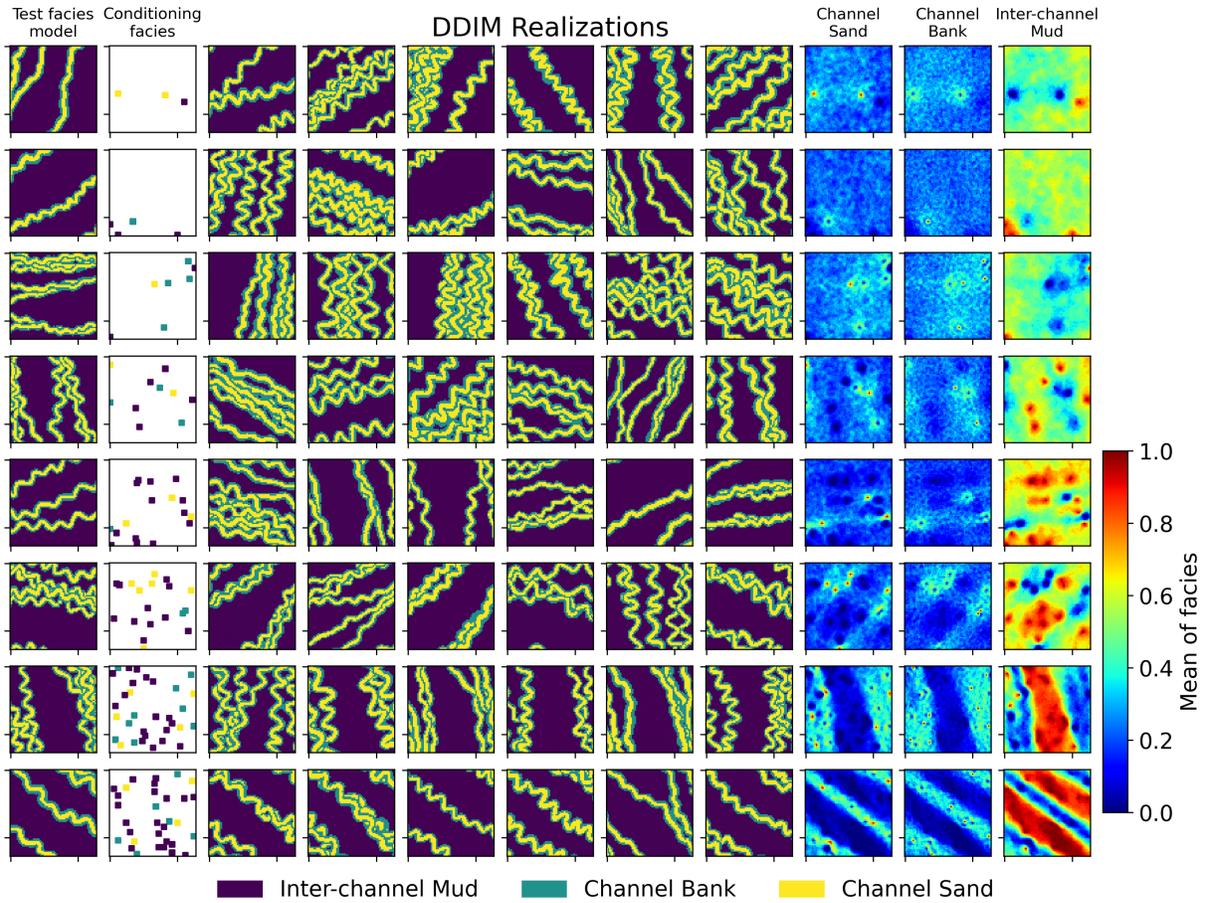

**Figure 17.** The generated realizations using conditional diffusion models with the different well facies data as constraints. Each row includes a facies model sampled from the test set, the corresponding well facies data, six generated realizations, and the frequency maps of the three facies types calculated from 100 realizations. The well facies points in the second column are enlarged from 1x1 pixels to 5x5 pixels for better visualization purposes in this figure.

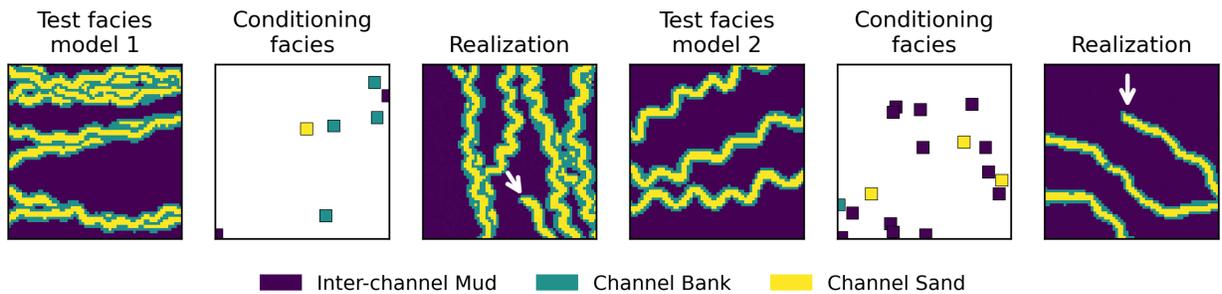

**Figure 18.** Two generated realizations exhibiting channel discontinuity artifacts. White

arrows indicate the locations where channel continuity is not preserved.

For the bottom two cases in Figure 17, we compare the referenced test facies model with the distributions of 100 random conditional realizations and 100 random unconditional realizations in a two-dimensional space using the combined MS-SWD and MDS method. The MDS plots are shown in Figure 19. Three conditional realizations with the largest, medium, and the lowest similarities (measured by the Euclidean distance in the two-dimensional MDS space) to the reference facies models are marked in yellow in the MDS plots. The light-blue points representing conditional models in Figure 19 cluster around the reference models (red crosses) relative to the unconditional model distributions (gray points), indicating that the input well-facies data impose an important constraint on the resulting geological patterns.

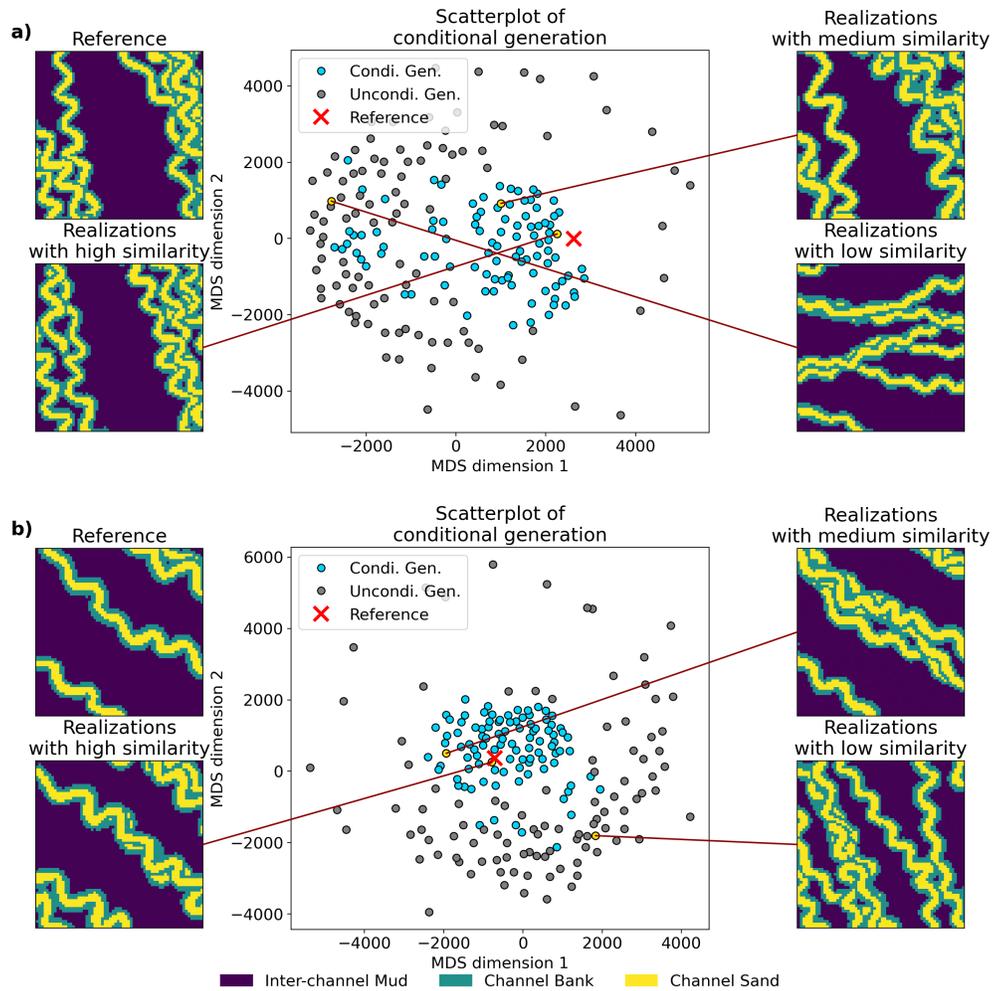

**Figure 19.** The MDS plots of the referenced test facies models, along with the conditional and unconditional generated realizations for two test cases. The three conditional realizations with the largest, medium, and the lowest similarities to the reference facies models are marked by yellow dots in the MDS plot and the respective conditional realizations and illustrated on the sides.

4.2.2 Point bar reservoir case

For the point bar reservoir case, the same procedures as in the channel case are followed to generate 100 realizations using the well facies data of eight random test facies models, as shown in Figure 20. The frequency maps of the four facies are calculated from 100 conditional realizations (the last four columns of Figure 20). With fewer input well facies points, the generated facies model realizations exhibit substantial diversity and geological realism, as evidenced by both the realizations themselves and the corresponding frequency maps (e.g., the top two rows in Figure 20). As the number of input well facies points increases, however, the realizations become progressively more constrained, and the overall diversity decreases. The reproduction accuracy of the input well facies data is 100% based on an evaluation of 800 conditional realizations across 8 cases in Figure 20. This demonstrates that the conditional generation process of DDIM can effectively constrain the geomodels to well facies data while maintaining the expected realism and diversity.

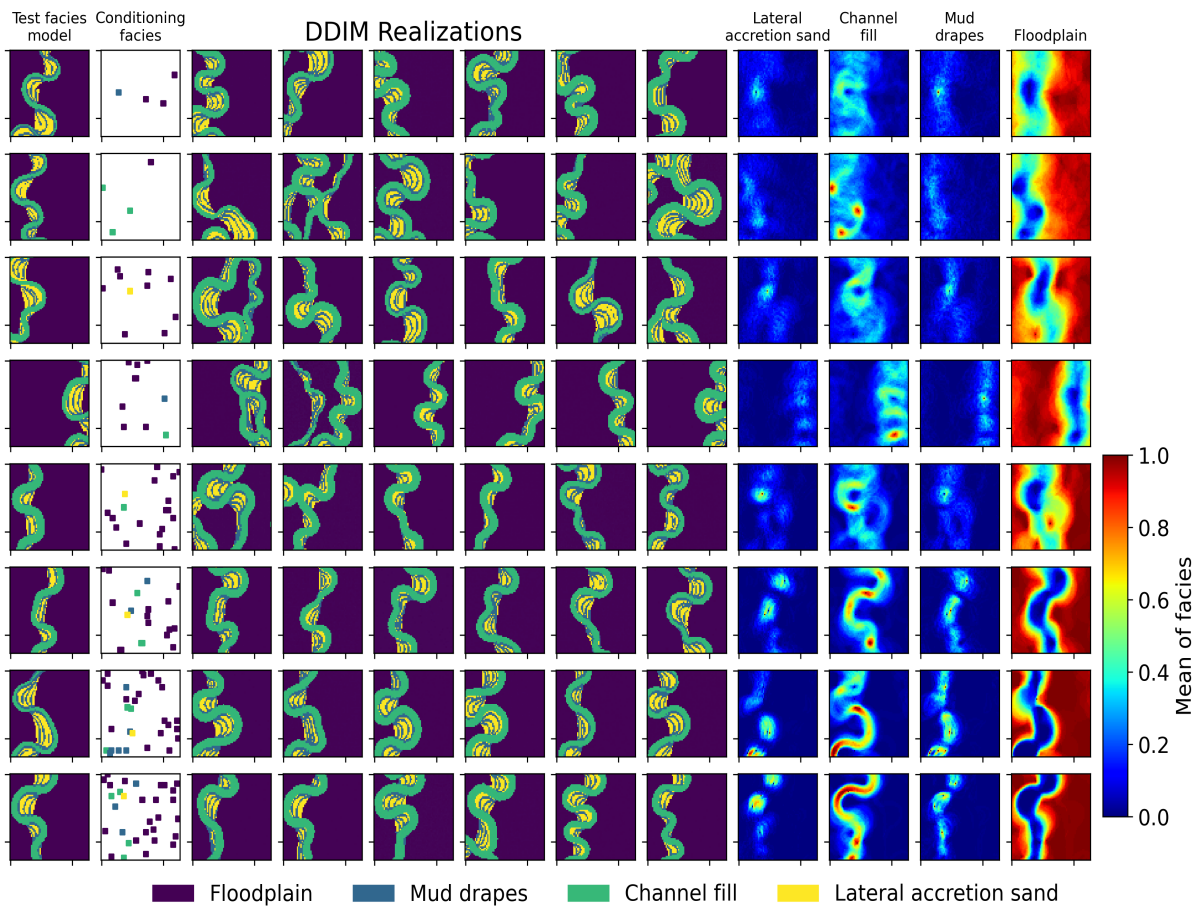

**Figure 20.** Conditional realizations generated from diffusion models with the different well facies data as constraints. Each row includes a geomodel from the test set, the corresponding well facies data, five randomly generated realizations, and frequency maps of the four facies types calculated from 100 realizations. The well facies points in the second column are enlarged from 1x1 pixels to 5x5 pixels for better visualization.

### 4.2.3 Three-dimensional point bar reservoir case

In this case, a 3D U-Net neural network is trained for denoising generative purposes, which is subsequently employed for conditional geomodelling following the DDIM inference workflow. Figure 21 presents the geomodelling results obtained using different test well facies data as conditioning input. Each row displays a test facies model, the corresponding input well facies, four randomly generated realizations, and the frequency and variance maps of the channel facies assemblage, including mud drapes, channel fill, and lateral accretion sand facies. The given well facies are reproduced with 100% accuracy in the examples of Figure 21. As shown in Figure 21, all generated realizations appear geologically realistic. With fewer input well facies data, the realizations exhibit substantial diversity, as reflected in both the realizations themselves and the corresponding frequency and variance maps in the top three rows. As the amount of input well facies data increases, however, the diversity of realizations gradually decreases, as evidenced by the progressively more certain features in the probability and variance maps from the fourth to the sixth rows.

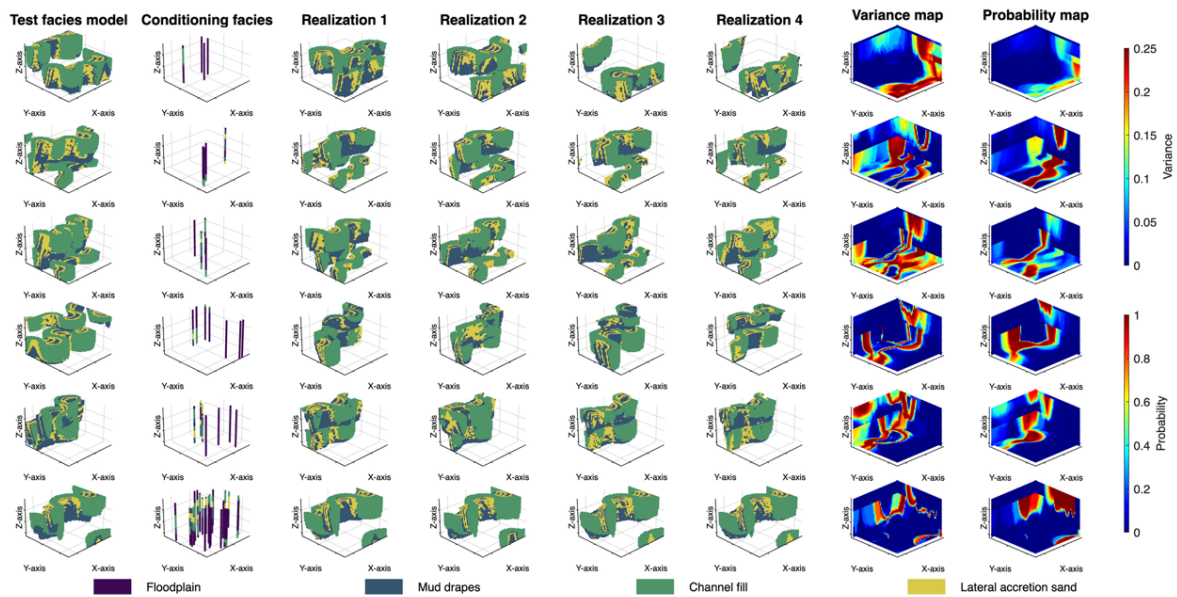

**Figure 21.** Conditional realizations generated from diffusion models with different known facies as constraints. Each row includes a geomodel from the test set, randomly assumed

known facies, four out of 100 realizations, and generated variance and probability maps of the river facies.

We observe that denoising diffusion models trained on the 3D dataset converge more slowly than those trained on the 2D dataset. Achieving strong performance takes approximately 32 hours of training on four A100 GPUs. We attribute this slower convergence to the increased difficulty of simultaneously honoring sparse well facies data and capturing complex 3D spatial geological patterns. The inference time for a single 3D facies model realization is approximately 2.08 s. To further improve generation efficiency during inference, latent diffusion models will be explored in future work.

## 5. Discussion

### 5.1 Comparing different training time steps for DDPM and DDIM

The total number of time steps in diffusion models may influence the generation realism, diversity, and training and inference time. To explore that influence, we investigate the generation results by setting the total time step number to 500, 1000, and 1500, respectively. The training times across different time-step settings are comparable in the channel-reservoir case, as demonstrated in Section 4. In contrast, the inference time increases proportionally with the time steps involved. Specifically, the inference times per realization are approximately 0.26, 0.53, and 0.8 seconds for 500, 1000, and 1500 time steps, respectively. Figure 22 shows randomly generated facies models produced by DDPM for the three time-step settings. The figure shows that the generated facies models are realistic and diverse across all three settings. However, when the results are projected into 2D plots with the combined method of MS-SWD and MDS (Figure 23), we see obvious distribution discrepancy between the 500 time step results and the test facies models, while the facies model distributions of the 1000 and 1500 time step results almost overlap with that of the test facies models. The reason may be that the difference between the hidden data distributions of adjacent time steps becomes smaller with increasing time steps, thus making the noise predictor easier to learn the mapping between those data distributions.

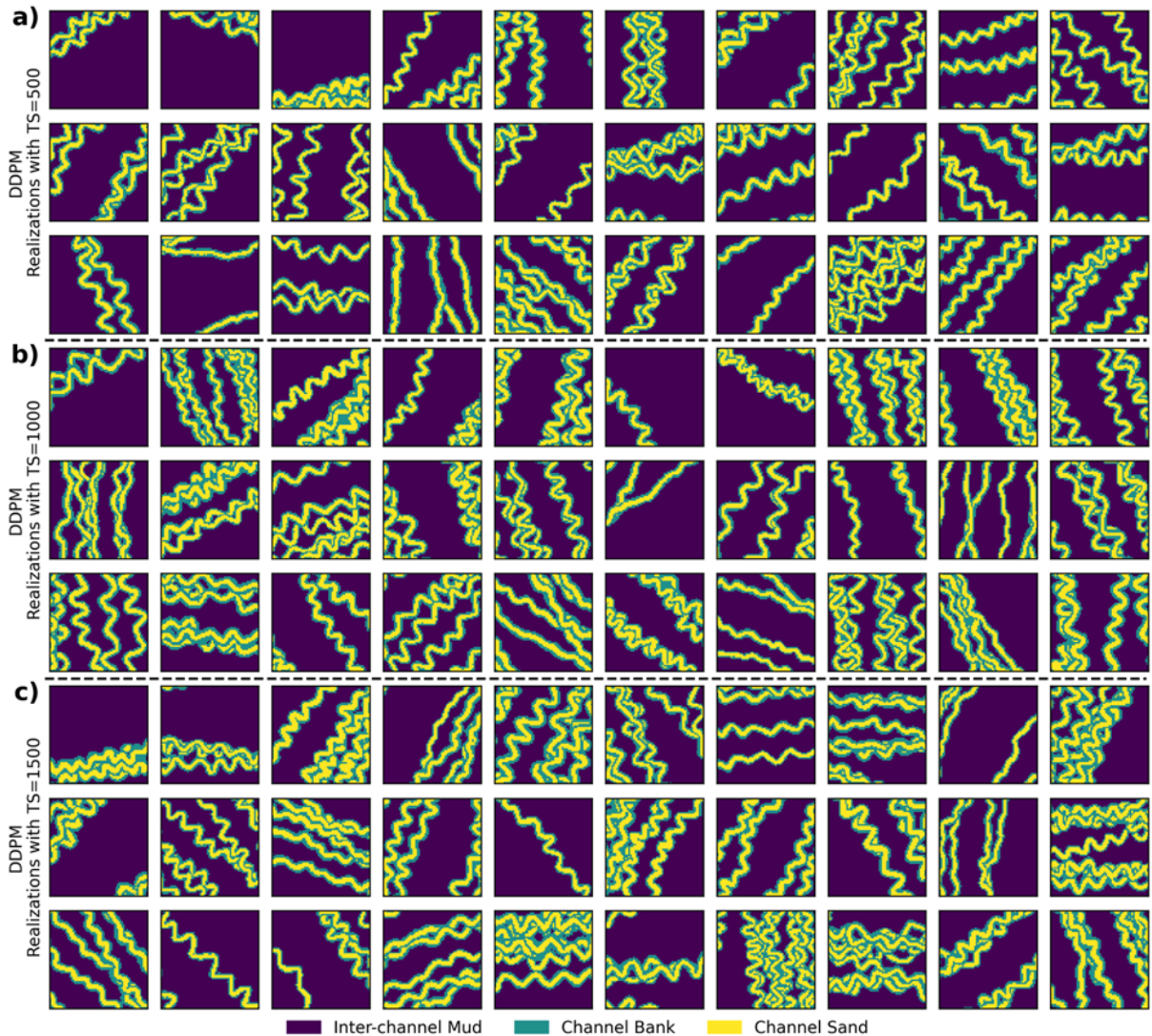

**Figure 22.** Random realizations are generated from DDPM when the total time steps are set as a) 500, b) 1000, and c) 1500. 'TS' in the labels denotes the total number of time steps used for training the diffusion model.

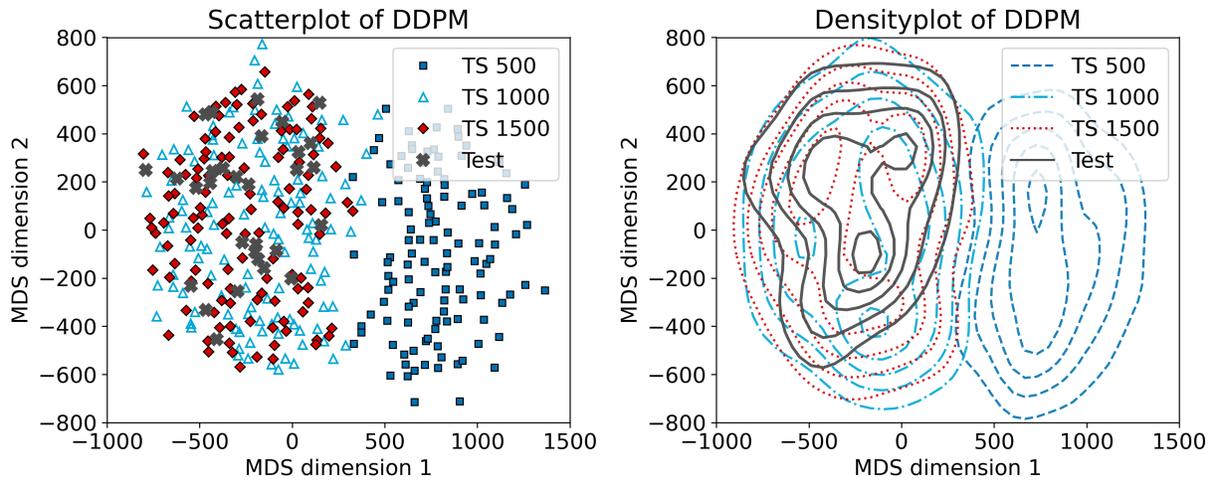

**Figure 23.** Multidimensional scaling plot of the test set and realizations generated using DDPM with 500, 1000, and 1500 time steps. 'TS' in the legends refers to the total time steps for diffusion model training.

We further investigate the generation results using DDIM for the three time step settings. The number of skipped time steps is set as 10, 20, and 30 for the three cases, respectively. Figure 24 shows random generation results for the three settings, and Figure 25 shows the MDS plots of generated results and test facies models. A mode collapse can be clearly seen from the 500 time step results, since all the generated facies models exhibit a high proportion of mud facies, which is inconsistent with the training/test dataset. In MDS plots, the 500 and the 1000 time step results are away from the test facies models, while the 1500 time step results match well with the test ones. One possible explanation is that increasing the number of training time steps may enhance the noise predictor network's performance, thereby reducing errors in the reverse diffusion process. This reduction in error could potentially facilitate the use of accelerated sampling algorithms, such as DDIM, by decreasing the likelihood of cumulative errors throughout the sampling process. Therefore, we recommend using relatively large total time steps for model training, even if we perform only a small number of iterations during inference.

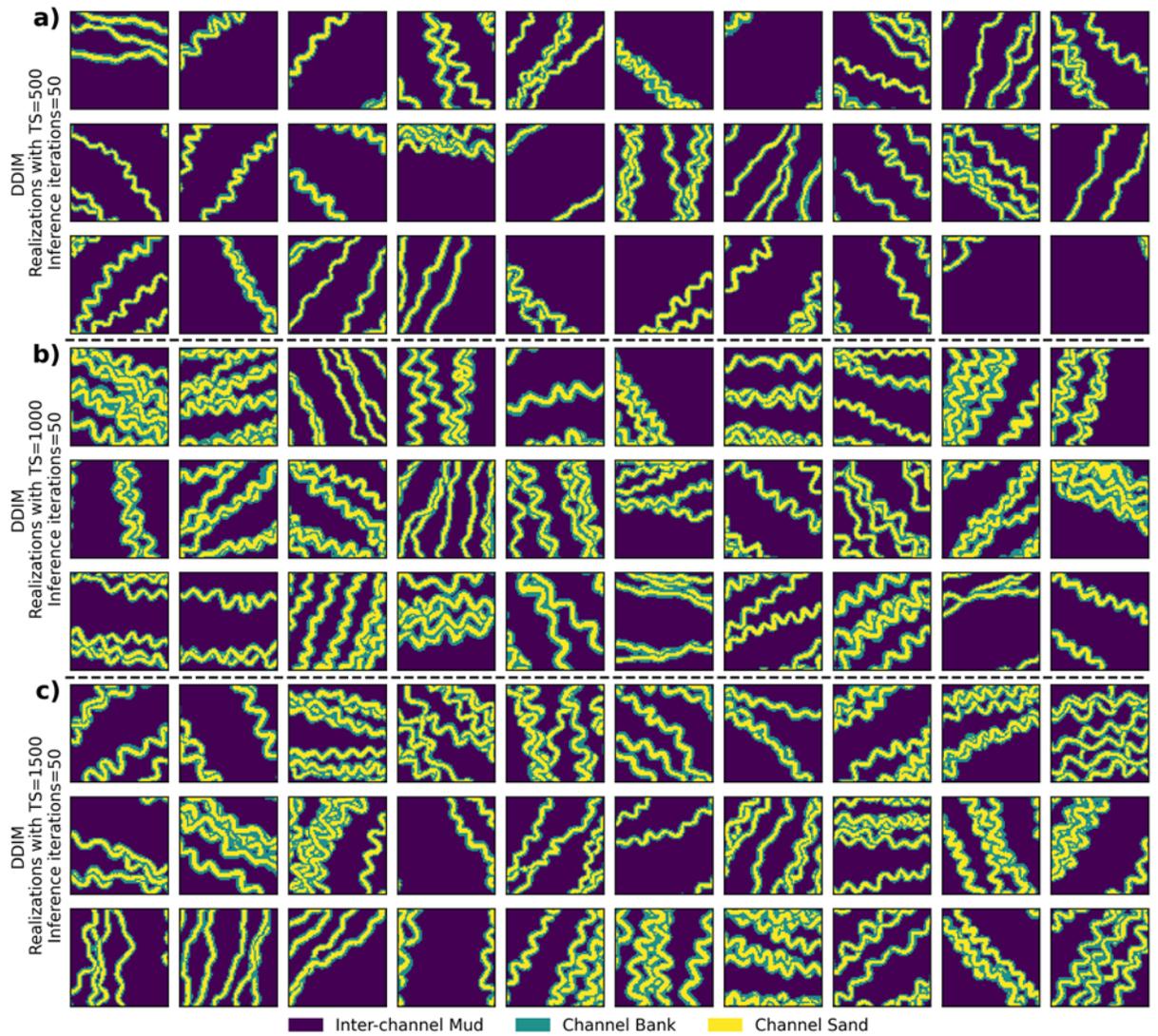

**Figure 24.** Random realizations are generated from DDIM when the total training time steps are set as a) 500, b) 1000, and c) 1500, but only 50 iterations are utilized at the inference stage.

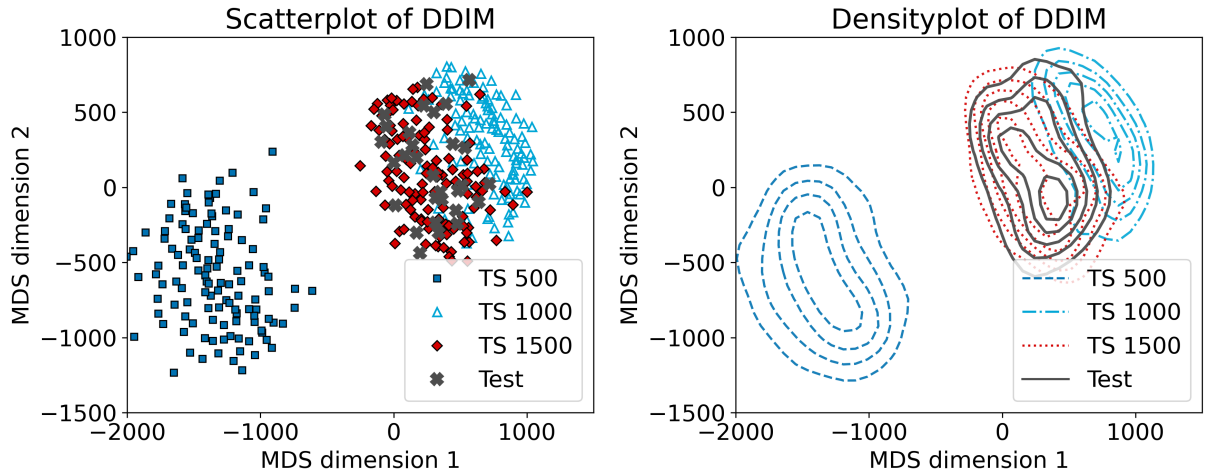

**Figure 25.** Multidimensional scaling plot of the test set and realizations generated using DDIM with 500, 1000, and 1500 time steps. 'TS' in the legends refers to the total time steps for diffusion model training.

5.2 Construct large geomodels from pre-trained denoising diffusion models

The trained diffusion models are not limited to generating models of only the same size as used during training, but can generate larger geomodels conditioned on denser well facies data. Because the fully convolutional neural network design is used in the noise predictor network, the trained model can be extended for geomodelling of arbitrarily large reservoirs as long as the size of the input Gaussian noise map and the output size of the geomodel remain the same. This fully convolutional property has also been exploited to scale GAN-based conditional geomodelling to larger 3D reservoirs (S. Song, Mukerji, et al., 2025; S. Song, Mukerji, Hou, et al., 2022). For the meandering 2D case, we applied the trained diffusion model (trained on 64 x 64 training set samples) to generate 128×128-pixel reservoirs, using 50 randomly selected well facies points as input. Note that during training, 3 to 32 conditioning points for the well facies are considered in each 64x64-cell. Figure 26 shows the input well facies points and three randomly generated facies models. The large-scale realizations preserve geological structures consistent with the training facies models and honor the conditioning wells exactly (100% honor rate at well locations). Figure 27 shows the large-size generated results conditioned on 80 well facies points for the point bar reservoir case. These three realizations are distinct, but both are conditioned on the given well facies. Figure 28 shows the large-size generated results with 128x128x64 cells, (trained on 48 x 48 x 32) which reproduce the input well facies data with 100% accuracy and are consistent with the geological patterns embedded in the 3D point bar facies models.

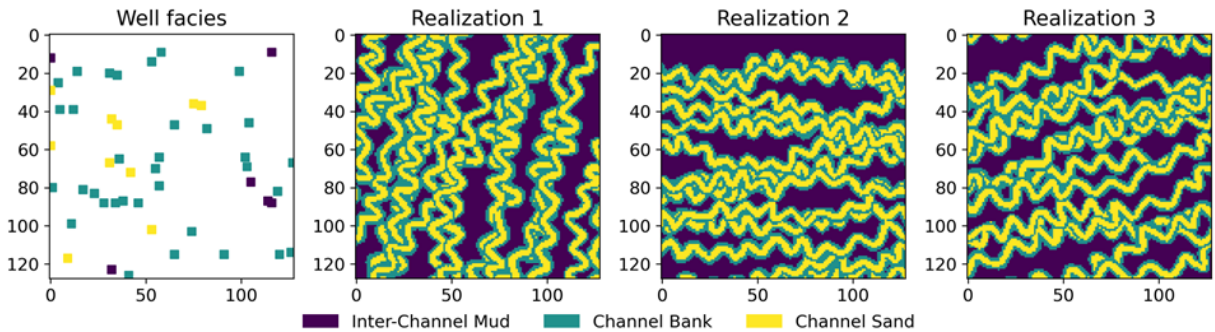

**Figure 26.** 2D meandering case. Conditioning wells and three realizations generated by the diffusion model under the same well facies constraints. Generated models are not limited to the size (64x64) used during training, but can generate larger model sizes (e.g. 128 x128).

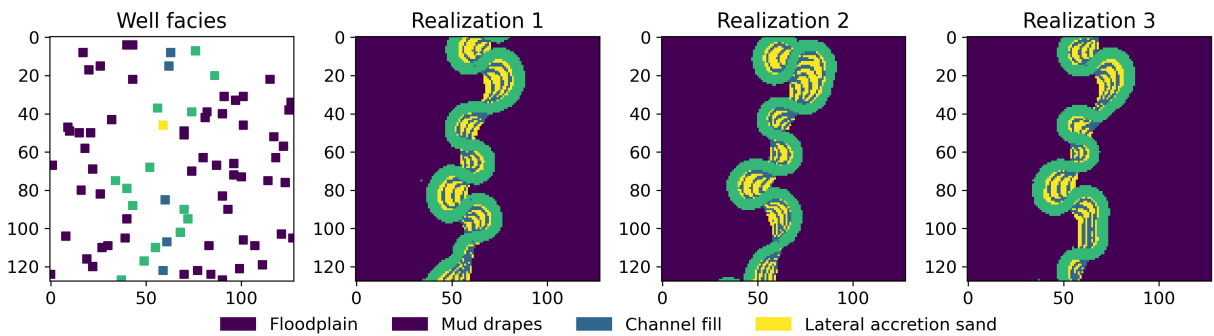

**Figure 27.** 2D point-bar case. Conditioning wells and three realizations generated by the diffusion model under the same well facies constraints. Generated models are not limited to the size (64x64) used during training, but can generate larger model sizes (e.g. 128 x128).

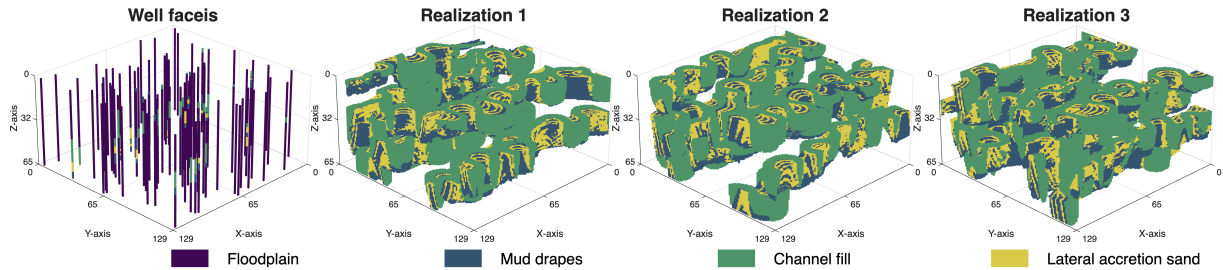

**Figure 28.** 3D point-bar case. Conditioning wells and three 3D realizations generated by the diffusion model under the same well facies constraints. Generated models are not limited to the size (48 x 48 x 32) used during training, but can generate larger model sizes (e.g. 128 x128 x 64).

## 6. Conclusions

We investigate denoising diffusion generative models for both unconditional sampling and geomodelling conditioned on well facies data, and evaluate their performance on three reservoir scenarios (two 2D cases and one 3D case). Across these scenarios, the trained diffusion model produces realistic, diverse facies realizations that match the test set across multiple statistics, including MDS distribution, facies proportions, variograms, and geometric features, demonstrating its ability to effectively capture realistic geological patterns. For conditional geomodelling, we propose a mask-based strategy that updates only inter-well regions while keeping the known well facies fixed during sampling, enabling hard conditioning without introducing additional loss reweighting. This framework enables uncertainty quantification by generating multiple conditioned realizations under hard well constraints. In addition, the trained diffusion models can be extended to conditional geomodelling of moderately large reservoirs while conditioning on a larger number of wells. Improving the training and inference efficiency of 3D diffusion models will be a focus of our future work.

**Acknowledgments**

We would like to acknowledge the sponsors of the Stanford Center for Earth Resources Forecasting (SCERF) for their support. Some of the computing for this project was performed on the Sherlock cluster. We would like to thank Stanford University and the Stanford Research Computing Center for providing computational resources and support that contributed to these research results.

**Use of AI Tools**

All content and figures in this manuscript were conceived, developed, and prepared by the authors. AI tools were used only for language editing and stylistic refinement. The authors take full responsibility for the accuracy, integrity, and presentation of the manuscript.

**Open Research**

The data and code used to train and run inference with the diffusion models are publicly available via Zenodo (Xu et al., 2026) and the project GitHub repository (https://github.com/minghuix98/DiffSim)

**Conflict of Interest Disclosure**

The authors declare there are no conflicts of interest for this manuscript.

**APPENDIX A: : Details of the Noise Predictor Network**

This section provides a concise overview of the architecture of the applied noise predictor network, following the discussion of Figure 2. Each encoding block consists of two consecutive residual blocks (denoted as Res Block), followed by a group normalization layer (denoted as Group norm) (Wu & He, 2018) and a linear attention layer (denoted as Linear Attention) (Shen et al., 2021), which is depicted in Figure A1a. A skip connection is established between the output of the second residual block and the output of the linear attention layer within the encoding blocks. Following the first and the second encoding blocks, downsampling layers are introduced to reduce both spatial dimensions by half, while simultaneously doubling the number of image channels, consistent with the standard U-Net architecture. The residual blocks employ a Wide ResNet block, as introduced by Zagoruyko (2016), in which the standard convolutional layer is replaced by a weight-standardized convolutional layer (denoted as WS Conv). The Res Block of Figure A1f includes two sub-blocks in which the time embedding information is integrated into the layer denoted as 'Scale & Shift' in the first block, which will be further described in the next paragraph. The SiLU within two sub-blocks both represent applying the Sigmoid Linear Unit (SiLU) activation function (Hendrycks & Gimpel, 2016). The third encoding block transitions into the hidden mid-block, which comprises a 3x3 convolutional layer (denoted as 3x3 Conv), two residual blocks, and an interposed attention layer (Vaswani et al., 2023), as illustrated in Figure A1c. The decoding block adopts the same architecture as the encoding block, with two skip connections linking the corresponding encoding and decoding blocks. The specific components are detailed in Figure A1b.

For temporal information processing, the time MLP block, as shown in Figure A1d, consists of a sinusoidal embedding for positional encoding (Vaswani et al., 2023), followed by a linear layer, a Gaussian Error Linear Unit (GELU; Hendrycks and Gimpel 2016) activation function, and another linear layer for output. The time embedding block, depicted in Figure A1e, incorporates a SiLU activation function and a linear layer for output. After being processed by the time MLP and time embedding layers, the specific time step is transformed into embedded time information, which serves as the input to the residual blocks, as indicated by the light blue arrows.

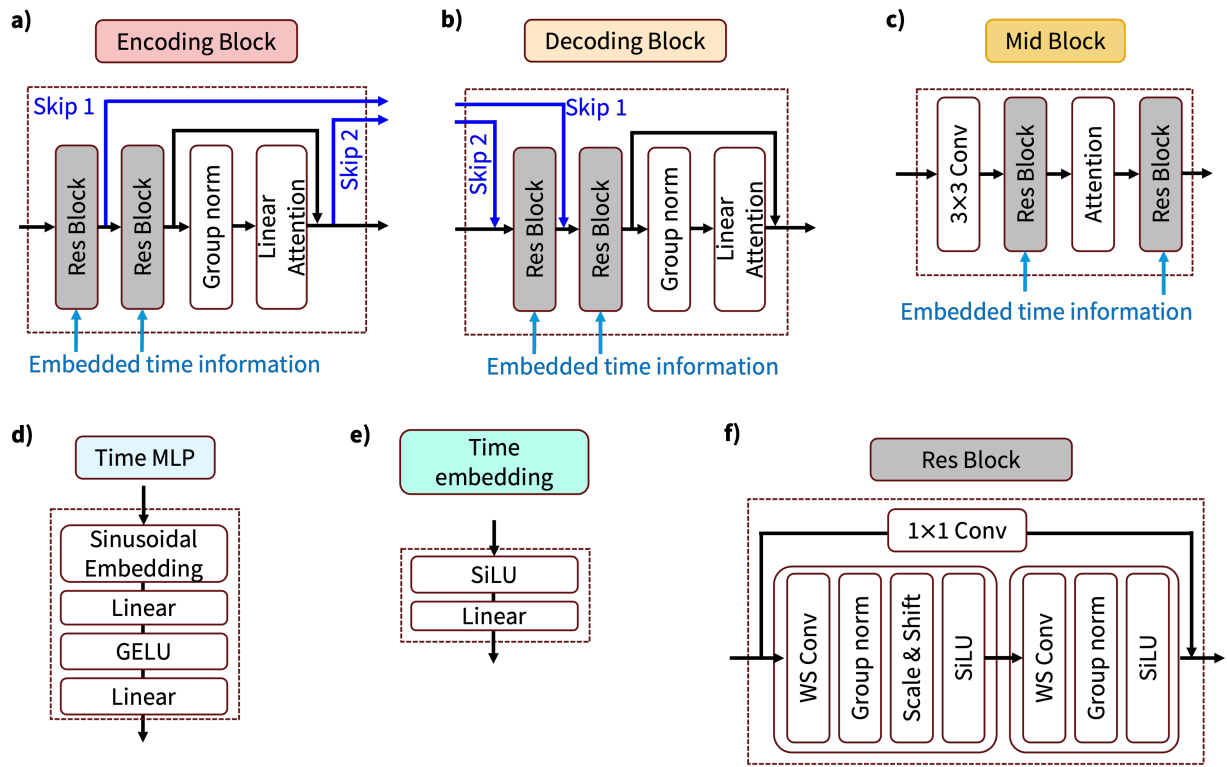

Figure A1. The specific components of different blocks: a) Encoding Block, b) Decoding Block, c) Mid Block, d) Time MLP, e) Time embedding, f) Res Block.

## References

Alqassab, H. M., Feng, M., Becker, J. A., Song, S., & Mukerji, T. (2024). MAGCS: Machine Assisted Geologic Carbon Storage (p. D021S044R001). Presented at the ADIPEC. https://doi.org/10.2118/222120-MS

Bond-Taylor, S., Leach, A., Long, Y., & Willcocks, C. G. (2022). Deep generative modelling: A comparative review of VAEs, GANs, normalizing flows, energy-based and autoregressive models. *IEEE Transactions on Pattern Analysis and Machine Intelligence*, *44*(11), 7327–7347. https://doi.org/10.1109/TPAMI.2021.3116668

Borg, I., & Groenen, P. J. F. (Eds.). (2005). MDS Models and Measures of Fit. In *Modern Multidimensional Scaling: Theory and Applications* (pp. 37–61). New York, NY: Springer New York. https://doi.org/10.1007/0-387-28981-X_3

Bosch, M., Mukerji, T., & Gonzalez, E. F. (2010). Seismic inversion for reservoir properties combining statistical rock physics and geostatistics: A review. *Geophysics*, *75*(5), 75A165-75A176. https://doi.org/10.1190/1.3478209

Chan, S., & Elsheikh, A. H. (2019). Parametric generation of conditional geological realizations using generative neural networks. *Computational Geosciences*, *23*(5), 925–952. https://doi.org/10.1007/s10596-019-09850-7

Chen, J., Huang, C.-K., Delgado, J. F., & Misra, S. (2023). Generating subsurface earth models using discrete representation learning and deep autoregressive network. *Computational Geosciences*, *27*(6), 955–974. https://doi.org/10.1007/s10596-023-10243-0

Cui, Z., Chen, Q., Liu, G., & Xun, L. (2024). SA-RelayGANs: A novel framework for the characterization of complex hydrological structures based on GANs and self-attention mechanismScaling. *Water Resources Research*, *60*(1), e2023WR035932. https://doi.org/10.1029/2023WR035932

Dhariwal, P., & Nichol, A. (2021). Diffusion models beat gans on image synthesis. *Advances in Neural Information Processing Systems*, *34*, 8780–8794.

Di Federico, G., & Durlofsky, L. J. (2024). Latent diffusion models for parameterization and data assimilation of facies-based geomodels. In *arXiv preprint arXiv:2406.14815*.

Di Federico, G., & Durlofsky, L. J. (2025). Three-Dimensional Latent Diffusion Models for Parameterizing and History Matching Facies Systems Under Hierarchical Uncertainty. *Mathematical Geosciences*, 1–33.

Esser, P., Kulal, S., Blattmann, A., Entezari, R., Müller, J., Saini, H., et al. (2024, March 5). Scaling rectified flow Transformers for high-resolution image synthesis. arXiv. Retrieved from http://arxiv.org/abs/2403.03206

Fuller, I., Reid, H., & Brierley, G. (2013). Methods in geomorphology: investigating river channel form. In *Treatise on geomorphology: Methods in geomorphology* (pp. 73–91). Elsevier.

Goodfellow, I., Pouget-Abadie, J., Mirza, M., Xu, B., Warde-Farley, D., Ozair, S., et al. (2014). Generative adversarial nets. In *Advances in neural information processing systems* (Vol. 27).

Hendrycks, D., & Gimpel, K. (2016). Gaussian error linear units (gelus). *arXiv Preprint arXiv:1606.08415*.

Ho, J., & Salimans, T. (2022, July 25). Classifier-free diffusion guidance. arXiv. Retrieved from http://arxiv.org/abs/2207.12598

Ho, J., Jain, A., & Abbeel, P. (2020). Denoising diffusion probabilistic models. *Advances in Neural Information Processing Systems*, *33*, 6840–6851.

Hu, X., Song, S., Hou, J., Yin, Y., Hou, M., & Azevedo, L. (2024). Stochastic modeling of thin mud drapes inside point bar reservoirs with ALLUVSIM-GANSim. *Water Resources Research*, *60*(6), e2023WR035989. https://doi.org/10.1029/2023WR035989

Karras, T., Aila, T., Laine, S., & Lehtinen, J. (2017). Progressive growing of GANs for improved quality, stability, and variation. *arXiv Preprint arXiv:1710.10196*.

Kingma, Diederik P, & Welling, M. (2013). Auto-encoding variational bayes. *arXiv Preprint arXiv:1312.6114*.

Kingma, Durk P, & Dhariwal, P. (2018). Glow: Generative flow with invertible 1x1 convolutions. *Advances in Neural Information Processing Systems*, *31*.

Kong, Z., Ping, W., Huang, J., Zhao, K., & Catanzaro, B. (2020). Diffwave: A versatile diffusion model for audio synthesis. In *arXiv preprint arXiv:2009.09761*.

Laloy, E., Hérault, R., Jacques, D., & Linde, N. (2018). Training-image based geostatistical inversion using a spatial generative adversarial neural network. *Water Resources Research*, *54*(1), 381–406. https://doi.org/10.1002/2017WR022148

Lee, D., Ovanger, O., Eidsvik, J., Aune, E., Skauvold, J., & Hauge, R. (2025). Latent diffusion model for conditional reservoir facies generation. *Computers & Geosciences*, *194*, 105750. https://doi.org/10.1016/j.cageo.2024.105750

Luo, X., Sun, J., Zhang, R., Chi, P., & Cui, R. (2024). A multi-condition denoising diffusion probabilistic model controls the reconstruction of 3D digital rocks. *Computers & Geosciences*, *184*, 105541. https://doi.org/10.1016/j.cageo.2024.105541

Mosser, L., Dubrule, O., & Blunt, M. J. (2020). Stochastic seismic waveform inversion using generative adversarial networks as a geological prior. *Mathematical Geosciences*, *52*(1), 53–79. https://doi.org/10.1007/s11004-019-09832-6

Nesvold, E., & Mukerji, T. (2021). Simulation of fluvial patterns with GANs trained on a data set of satellite imagery. *Water Resources Research*, *57*(5), e2019WR025787. https://doi.org/10.1029/2019WR025787

Oliver, M. A., & Webster, R. (2014). A tutorial guide to geostatistics: Computing and modelling variograms and kriging. *CATENA*, *113*, 56–69. https://doi.org/10.1016/j.catena.2013.09.006

Ramesh, A., Dhariwal, P., Nichol, A., Chu, C., & Chen, M. (2022). Hierarchical text-conditional image generation with clip latents. *arXiv Preprint arXiv:2204.06125*, *1*(2), 3.

Rombach, R., Blattmann, A., Lorenz, D., Esser, P., & Ommer, B. (2022). High-resolution image synthesis with latent diffusion models. In *2022 IEEE/CVF Conference on Computer Vision and Pattern Recognition (CVPR)* (pp. 10674–10685). IEEE. https://doi.org/10.1109/CVPR52688.2022.01042

Saharia, C., Chan, W., Chang, H., Lee, C., Ho, J., Salimans, T., et al. (2022). Palette: image-to-image diffusion models. In *ACM SIGGRAPH 2022 Conference Proceedings*. Vancouver, BC, Canada: Association for Computing Machinery. https://doi.org/10.1145/3528233.3530757

Saxena, D., & Cao, J. (2021). Generative adversarial networks (GANs): challenges, solutions, and future directions. *ACM Comput. Surv.*, *54*(3). https://doi.org/10.1145/3446374

Shen, Z., Zhang, M., Zhao, H., Yi, S., & Li, H. (2021). Efficient attention: Attention with linear complexities. In *Proceedings of the IEEE/CVF winter conference on applications of computer vision* (pp. 3531–3539).

Song, J., Meng, C., & Ermon, S. (2022, October 5). Denoising diffusion implicit models. arXiv. Retrieved from http://arxiv.org/abs/2010.02502

Song, S., Mukerji, T., & Hou, J. (2021a). GANSim: Conditional facies simulation using an improved progressive growing of generative adversarial networks (GANs). *Mathematical Geosciences*, *53*(7), 1413–1444. https://doi.org/10.1007/s11004-021-09934-0

Song, S., Mukerji, T., & Hou, J. (2021b). Geological facies modeling based on progressive growing of generative adversarial networks (GANs). *Computational Geosciences*, *25*(3), 1251–1273. https://doi.org/10.1007/s10596-021-10059-w

Song, S., Mukerji, T., & Hou, J. (2022). Bridging the gap between geophysics and geology with generative adversarial networks. *IEEE Transactions on Geoscience and Remote Sensing*, *60*, 1–11. https://doi.org/10.1109/TGRS.2021.3066975

Song, S., Mukerji, T., Hou, J., Zhang, D., & Lyu, X. (2022). GANSim-3D for conditional geomodeling: theory and field application. *Water Resources Research*, *58*(7), e2021WR031865. https://doi.org/10.1029/2021WR031865

Song, S., Zhang, D., Mukerji, T., & Wang, N. (2023). GANSim-surrogate: An integrated framework for stochastic conditional geomodelling. *Journal of Hydrology*, *620*, 129493. https://doi.org/10.1016/j.jhydrol.2023.129493

Song, S., Huang, J., & Mukerji, T. (2025). Generative geomodelling: Deep Learning vs. Geostatistics. https://doi.org/10.31223/X5B732

Song, S., Mukerji, T., Scheidt, C., Alqassab, H., & Feng, M. (2025). Geomodelling of multi-scenario non-stationary reservoirs with enhanced GANSim. https://doi.org/10.31223/X5Z73N

Song, Y., & Ermon, S. (2019). Generative modeling by estimating gradients of the data distribution. In *Advances in neural information processing systems* (Vol. 32).

Song, Y., Sohl-Dickstein, J., Kingma, D. P., Kumar, A., Ermon, S., & Poole, B. (2020). Score-based generative modeling through stochastic differential equations. *arXiv Preprint arXiv:2011.13456*.

Strebelle, S. (2002). Conditional simulation of complex geological structures using multiple-point statistics. *Mathematical Geology*, *34*(1), 1–21. https://doi.org/10.1023/A:1014009426274

Tahmasebi, P. (2018). Multiple point statistics: A review. In B. S. Daya Sagar, Q. Cheng, & F. Agterberg (Eds.), *Handbook of Mathematical Geosciences: Fifty Years of IAMG* (pp. 613–643). Cham: Springer International Publishing. https://doi.org/10.1007/978-3-319-78999-6_30

Vaswani, A., Shazeer, N., Parmar, N., Uszkoreit, J., Jones, L., Gomez, A. N., et al. (2023, August 1). Attention is all you need. arXiv. Retrieved from http://arxiv.org/abs/1706.03762

Voleti, V., Yao, C.-H., Boss, M., Letts, A., Pankratz, D., Tochilkin, D., et al. (2024). SV3D: Novel multi-view synthesis and 3D generation from a single image using latent video diffusion. *arXiv Preprint arXiv:2403.12008*.

Wang, F., Huang, X., & Alkhalifah, T. (2024). Controllable seismic velocity synthesis using generative diffusion models. In *arXiv preprint arXiv:2402.06277*.

Wang, T.-C., Liu, M.-Y., Zhu, J.-Y., Liu, G., Tao, A., Kautz, J., & Catanzaro, B. (2018). Video-to-video synthesis. *arXiv Preprint arXiv:1808.06601*.

Wu, Y., & He, K. (2018). Group normalization. In *Proceedings of the European conference on computer vision (ECCV)* (pp. 3–19).

Xing, Z., Feng, Q., Chen, H., Dai, Q., Hu, H., Xu, H., et al. (2023). A survey on video diffusion models. *arXiv Preprint arXiv:2310.10647*.

Xu, M., Song, S., & Mukerji, T. (2026). DiffSIM: Unconditional and conditional facies simulation based on denoising diffusion generative models [Data set]. Zenodo. https://doi.org/10.5281/ZENODO.18904773

Yang, L., Zhang, Z., Song, Y., Hong, S., Xu, R., Zhao, Y., et al. (2024). Diffusion models: A comprehensive survey of methods and applications. *ACM Computing Surveys*, *56*(4), 1–39. https://doi.org/10.1145/3626235

Zagoruyko, S. (2016). Wide residual networks. *arXiv Preprint arXiv:1605.07146*.

Zhang, T.-F., Tilke, P., Dupont, E., Zhu, L.-C., Liang, L., & Bailey, W. (2019). Generating geologically realistic 3D reservoir facies models using deep learning of sedimentary architecture with generative adversarial networks. *Petroleum Science*, *16*(3), 541–549. https://doi.org/10.1007/s12182-019-0328-4